\documentclass[12pt, preprint]{aastex}
\usepackage{epsf,color,cancel,algorithmic,amsmath,subfigure}

\newcommand{\sfrac}[2]{\mathchoice
  {\kern0em\raise.5ex\hbox{\the\scriptfont0 #1}\kern-.15em/
    \kern-.15em\lower.25ex\hbox{\the\scriptfont0 #2}}
  {\kern0em\raise.5ex\hbox{\the\scriptfont0 #1}\kern-.15em/
    \kern-.15em\lower.25ex\hbox{\the\scriptfont0 #2}}
  {\kern0em\raise.5ex\hbox{\the\scriptscriptfont0 #1}\kern-.2em/
    \kern-.15em\lower.25ex\hbox{\the\scriptscriptfont0 #2}} {#1\!/#2}}

\usepackage{multirow}

\newcommand{\isot}[2]{$^{#2}\mathrm{#1}$}
\newcommand{\isotm}[2]{{}^{#2}\mathrm{#1}}

\newcommand{\rhob}{\ensuremath{\rho_\mathrm{base}}}
\newcommand{\Tb}{\ensuremath{T_\mathrm{base}}}
\newcommand{\gcc}{\mathrm{g~cm^{-3} }}

\newcommand{\pg}{\ensuremath{\left(\mathrm{p},\gamma\right)}}
\newcommand{\gp}{\ensuremath{\left(\gamma,\mathrm{p}\right)}}
\newcommand{\ap}{\ensuremath{\left(\alpha,\mathrm{p}\right)}}

\newcommand{\ag}{\ensuremath{\left(\alpha,\gamma\right)}}
\newcommand{\weak}{\ensuremath{\left(,\mathrm{e}^+\nu\right)}}

\begin{document}
\title{Multidimensional Modeling of Type I X-ray
  Bursts. II. Two-Dimensional Convection in a Mixed H/He Accretor}

\shorttitle{Multidimensional Modeling of Type I X-ray Bursts II.}
\shortauthors{Malone et al.}

\author{C.~M.~Malone\altaffilmark{1,2},
        M.~Zingale\altaffilmark{2},
        A.~Nonaka\altaffilmark{3},
        A.~S.~Almgren\altaffilmark{3},
        J.~B.~Bell\altaffilmark{3}}
\email{malone@ucolick.org}

\altaffiltext{1}{Department of Astronomy \& Astrophysics,
                 The University of California, Santa Cruz,
                 Santa Cruz, CA 95064, USA}

\altaffiltext{2}{Dept. of Physics \& Astronomy,
                 Stony Brook University,
		 Stony Brook, NY 11794-3800}

\altaffiltext{3}{Center for Computational Sciences and Engineering,
                 Lawrence Berkeley National Laboratory,
                 Berkeley, CA 94720}

\begin{abstract}
Type I X-ray Bursts (XRBs) are thermonuclear explosions of accreted
material on the surfaces of a neutron stars in low mass X-ray
binaries.  Prior to the ignition of a subsonic burning front, runaway
burning at the base of the accreted layer drives convection that mixes
fuel and heavy-element ashes.  In this second paper in a series, we
explore the behavior of this low Mach number convection in mixed
hydrogen/helium layers on the surface of a neutron star using
two-dimensional simulations with the Maestro code.  Maestro takes
advantage of the 
highly subsonic flow field by filtering dynamically unimportant sound
waves while retaining local compressibility effects, such as those due
to stratification and energy release from nuclear reactions.  In these
preliminary calculations, we find that the rp-process approximate
network creates a convective region that is split into two layers.
While this splitting appears artificial due to the approximations of
the network regarding nuclear flow out of the breakout reaction
\isot{Ne}{18}\ap\isot{Na}{21}, these calculations hint at further
simplifications and improvements of the burning treatement for use in
subsequent calculations in three dimensions for a future paper.
\end{abstract}

\section{Introduction}\label{Sec:Introduction}

An accreting neutron star can only build up a thin ($\sim 10$~m)
surface layer of H/He before the immense gravitational acceleration
compresses this fuel to the point of ignition.  The ensuing
thermonuclear runaway is short lived (10--100~s) but releases an
enormous flux of X-rays (total energy $\sim10^{40}$ ergs)---a
transient event we detect and classify as a Type I X-ray Burst (XRB)
(see
\citealt{lewin81,lewin93,BILDSTEN00,strohmayerbildsten2003,intZand2011}
for reviews).  Once the explosion subsides, the accretion builds up a
fresh layer of fuel in a matter of hours to days, and a new outburst
occurs.  An XRB lightcurve shows a sharp rise---about an order of
magnitude increase during $\sim 1~\mathrm{s}$ --- in the X-ray
luminosity followed by an extended ($\sim 10~\mathrm{s}$) decay.

Some ultra-compact systems are thought to accrete pure \isot{He}{4}
(4U 1820-30, for example; \citealt{CUMMING_03}).  The most common
systems, however, likely accrete a mixture of H/He from an evolved
companion star (see, for example, the compilation of bursts in
\citealt{galloway:2008}). Depending on the local accretion rate, the
\isot{H}{1} accreted in these systems may either 1) burn stably to
form a pure \isot{He}{4} layer, which then experiences a thin-shell
instability resulting in an outburst, or 2) become unstable itself in
the presence of helium resulting in a mixed outburst
\citep{fujimoto1981,FL87_XRB,CUMMING_BILDSTEN_00,BILDSTEN00}.  Mixed
bursts typically have longer lightcurves due to the waiting points in
the weak nuclear reactions (see \citealt{strohmayerbildsten2003} for
an overview).

Mixed H/He XRBs are important sites of explosive hydrogen burning via
the rp-process~\citep{wallacewoosley:1981,rpprocess,parikh:2014}.  The
nuclear physics of the rp-process nuclei is a focus of the
U.S. Department of Energy proposed Facility for Rare Isotope Beams.
Understanding the conditions that exist in XRBs is critical to
accurately modeling the nucleosynthesis, which may then alter the
lightcurve.  Furthermore, the subset of XRBs exhibiting so-called
Photospheric Radius Expansion (PRE) burst phenomena --- whereby the
burst's luminosity is large enough to lift the photosphere to larger
radii (lower effective temperature) before settling back down to the
neutron star surface---can yield information about neutron star masses
and radii (see for
example~\citealt{bhattacharyya:2010,ozel:2010,steiner:2010}).

Most of our theoretical understanding on XRBs comes from one-dimensional studies
with stellar evolution codes, assuming spherical symmetry.  These one-dimensional
calculations are able to roughly reproduce the observed energies,
durations, and recurrence timescales for XRBs
\citep{taam1980,TAAM_ETAL93,taamwoosleylamb1996,woosley-xrb,fisker:2008}.
Due to their one-dimensional nature, these simulations can use larger reaction
networks than multi-dimensional studies to predict the nucleosynthetic
yields from advanced burning stages, like the rp-process, and explore
the nucleosynthesis in
detail~\citep{schatz:rp1999,rpprocess,woosley-xrb,fisker:2008,jose:2010,parikh:2013}.
However, the one-dimensional nature prevents the simulations from directly
modeling the convection, and simplified models like mixing length
theory (see, for example~\citealt{KipWei}) are needed.  Recent
multi-dimensional simulations have questioned the validity of mixing
length theory, and emphasized the role of
turbulence~\citep{meakin:2007,arnett:2009}.  If the convection is not
modeled properly, then the wrong temperature/pressure history for a
fluid element will be obtained, affecting the nucleosynthesis and
lightcurve.

Simulations of the vertical structure of reacting flow on neutron
stars are rare.  Several models of detonations
\citep{FRY_WOOS_DETONATION_82,ZINGALE_ETAL01,simonenko:2012} have been
done, but these sample density regimes that are not typical of an XRB.
\citet{Lin:2006} used a low Mach number algorithm to model pure helium
bursts in two-dimensions, following the rise in temperature and
watching the development of convection.  \citet{cavecchi:2012} used a
simplified hydrodynamic model (the vertical direction was treated as
hydrostatic) to model flame propagation across the neutron star on
length and timescales appropriate to an XRB, but the numerical
technique does not allow for a detailed understanding of the dynamics
at the front, including mixing and turbulence.  It is also important
to understand whether the convection can bring ashes up to the
photosphere~\citep{intZand:2010,bhattacharyya:2010}, altering our
interpretation of the radiation.

In paper~I~\citep{xrb} we explored the convective dynamics of a pure
helium XRB using our low Mach number simulation code, Maestro.  Our
results differed from those of \citet{Lin:2006} in that we found that
a much higher resolution is needed to resolve the He burning peak and
to properly capture the convective dynamics.  Here we extend the low
Mach number methodology to the case of mixed H/He bursts with an
extended network that captures the hydrogen burning.  Our ultimate
goal in this series of papers is to evolve the convective region to
the point where we can see a nonlinear rise in the temperature, and to
assess how the convection impacts the nucleosynthesis.

\section{Numerical Method}\label{Sec:Numerical Method}

We use the Maestro algorithm as described in
\citet{MAESTRO:Multilevel}.  Maestro solves the equations of low Mach
number hydrodynamics applicable to hydrostatic (HSE) stellar
flows---that is, fluid quantities, e.g. density $\rho({\bf x},t)$, are
decomposed into radial background, HSE components, $\rho_0(r,t)$, and
perturbational components, $\rho'({\bf x},t)$, that govern the
dynamics (see \citet{MAESTRO:Multilevel} for details).  The background
pressure, $p_0(r,t)$, governs the thermodynamics, and is used in place
of the total pressure, $p = p_0 + p'$, everywhere except in the
momentum equation---an approximation valid for low Mach number flow up
to $\mathcal{O}(M^2)$.  A key feature of this equation set is that
soundwaves are filtered from the system, enabling long timescale
evolution of highly subsonic flows.  This manifests itself through an
elliptic constraint on the velocity field that captures the effects of
compressibility on a fluid element due to localized heating and
stratification.  The overall algorithm consists of an advection step,
using the piecewise parabolic method, a projection step that enforces
the velocity constraint, and reactions, coupled in via Strang
splitting for second-order accuracy~\citep{ABNZ:III}.  Maestro is
publicly available.\footnote{Maestro can be found here: {\tt
    http://bender.astro.sunysb.edu/Maestro/ }}

For all simulations in this paper, we use the formulation of Maestro
that couples the enthalpy equation into the solution---that is, when
needed, the temperature is derived from density ($\rho$), enthalpy
($h$), and composition ($X_k$).  In computing the fluxes through the
interfaces of the cells, we need time-centered interface values of the
state variables.  There are a variety of quantities we can predict to
the edges to construct these interface states.  Experience has shown
that in the presence of steep composition gradients, the advection
algorithm is more stable when we predict the full state to the
interfaces rather than the perturbational fluid quantites.  In the
present work, we predict $\rho$ and $X_k$ to interfaces and compute
the interface state as $(\rho X_k)_\mathrm{edge} = \rho_\mathrm{edge}
X_{k,\mathrm{edge}}$, instead of predicting $\rho'$ and averaging
$\rho_0$ to interfaces and constructing the edge state as $(\rho_0 +
\rho'_\mathrm{edge}) X_{k,\mathrm{edge}}$.  For the enthalpy state, we
now predict $T$ to the interfaces, as this is the most sensitive
thermodynamic quantity.  We then use the equation of state to compute
$h$ on the interfaces and compute the enthalpy state as $(\rho
h)_\mathrm{edge} = \rho_\mathrm{edge} h(T_\mathrm{edge})$.  This
differs from the previous method of predicting $(\rho h)'$ to edges,
averaging $(\rho h)_0$ to edges, and then constructing the edge state
as $(\rho h)_0 + (\rho h)'_\mathrm{edge}$.  This change is more in
line with the original reconstruction described in \citet{ABRZ:I}, and
that used in \citet{subchandra}, but continues to be a subject of
algorithmic investigation.

Additionally, all the models we run here include the additional term
in the momentum equation identified by \citet{kleinpauluis} and
\citet{vasil:2013} that improves the energy conservation in the low
Mach number limit as well as the treatment of gravity waves (we
explore this for a variety of problems in \citealt{maestroenergy}).
We note, however, that this additional term has little effect in a
convective layer, but we include it for completeness.

\subsection{Microphysics}

We use an approximate network containing 10 species, based on the
description of \citet{wallacewoosley:1981}, Appendix C.  This network
approximates the hot CNO burning, triple-$\alpha$, plus rp-process
breakout and burning up through \isot{Ni}{56}.  For details, see the
\citet{wallacewoosley:1981} paper, but below we describe some of the
features of the approximate network.  We note that our implementation
follows that description faithfully, but uses updated ReacLib
\citep{ReacLib} rates where they exist; Table \ref{table:rates} lists
the particular rates used at some stage in the network, along with the
version identifier from ReacLib.  The stiff system of ODEs is
integrated using the VODE package~\citep{vode}.

\begin{figure}
\centering
\includegraphics{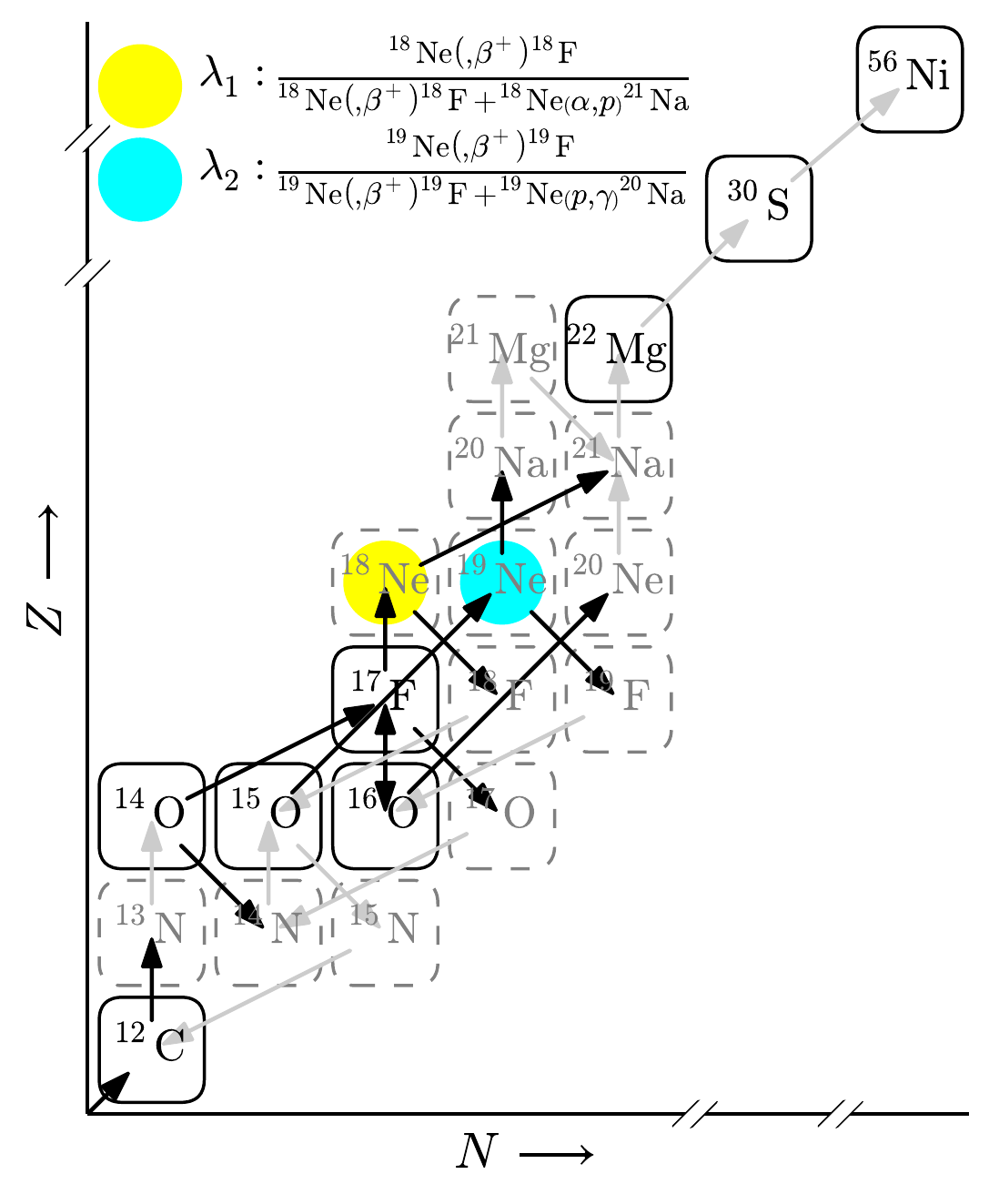}
\caption{\label{fig:rprox_schematic} Schematic of the rp-process
  approximate network.  Isotopes labelled in black are those used in
  the network, in addition to \isot{H}{1} and \isot{He}{4}; light gray
  isotopes with dashed boxes are those implicit in some of the
  reaction sequences.  Likewise, black lines denote explicit reactions
  calculated in the network, whereas gray lines mark where
  approximations are made.  The two circles mark important
  branching ratios for breaking out of the HCNO cycle.  See the text
  for more details.}
\end{figure}

Figure \ref{fig:rprox_schematic} shows a schematic of the rp-process
network outlined in \citet{wallacewoosley:1981} and used in this
paper.  The isotopes labelled in black are those explicitly included
in the network, in addition to \isot{H}{1} and \isot{He}{4}; likewise,
black arrows indicate reaction rates explicitly calculated, whereas
gray arrows denote approximations.  For example, in the HCNO cycle the
sequence of reactions
$\isotm{O}{14}\left(,\beta^+\right)\isotm{N}{14}\left(p,\gamma\right)\isotm{O}{15}$
is restricted by the slowest rate in the sequence, the
$\beta^+$-decay.  In the approximate network, \isot{O}{14} (plus a
proton) is converted {\em directly} to \isot{O}{15} at a rate given by
the $\beta^+$-decay.  The two colored circles in Figure
\ref{fig:rprox_schematic} denote locations where the nuclear flow can
break out of the HCNO cycle and start up the rp-process path to
heavier nuclei.  Each of these breakout points involve competition
between a $\beta^+$-decay of an isotope of Ne and an
$\left(\alpha,p\right)$ or $\left(p,\gamma\right)$ reaction with
branching ratios given by $\lambda_1$ or $\lambda_2$, respectively.
For example, the chain of reactions
$\isotm{O}{15}\left(\alpha,\gamma\right)\isotm{Ne}{19}\left(p,\gamma\right)\isotm{Na}{20}\left(p,\gamma\right)\isotm{Mg}{21}\left(,\beta^+\right)\isotm{Na}{21}\left(p,\gamma\right)\isotm{Mg}{22}$
is approximated as turning \isot{O}{15} (plus an $\alpha$ and 3
protons) directly into \isot{Mg}{22} at a rate goverened by the
$\alpha$-capture on \isot{O}{15} multiplied by the fraction
$(1-\lambda_2)$ of \isot{Ne}{19} isotopes that proton-capture before
they $\beta^+$-decay.  At the high temperatures and densities
experienced during an XRB, many of the $\beta^+$-decay waiting points
of the traditional rp-process are bypassed via $\left(p,\gamma\right)$
and/or $\left(\alpha,p\right)$ reactions.  For the flow in the
rp-process up through \isot{Ni}{56}, the approximate network accounts
for these bypassing reactions by assuming {\em all} reactions pass
through the fastest path.  For example, converting \isot{Mg}{22} to
\isot{S}{30} can proceed through two main channels: 1) a series of
eight $p$-captures and several $\beta^+$-decays, or 2) two
$\left(\alpha,p\right)$ and two $\left(p,\gamma\right)$ reactions.
The faster of the two paths will be used by the network.  However,
each individual path is limited by its slowest rate.  For the first
path---the $\beta^+$-decay path---the limiting factor is the mean
lifetime of the $\beta^+$-decays; the second path is limited by the
slower of the two $\alpha$-capture rates, namely
\isot{Si}{26}$\left(\alpha,p\right)$\isot{P}{29}.  Likewise, burning
\isot{S}{30} to \isot{Ni}{56} in the approximate network is restricted
by a typical $\beta^+$-decay timescale or $\alpha$-capture on
\isot{Ti}{44}.  Finally, \isot{He}{4} burning via the triple-$\alpha$
reaction enters the diagram from the bottom left corner.

A general stellar equation of state with ideal ions, arbitrarily
degenerate/relativistic electrons, and radiation is used
\citep{timmes_swesty:2000}.  Paper~I showed that including thermal
diffusion did not significantly alter the average properties of the
convection or the maximum temperature reached during the simulation.
Consequently, we neglect thermal diffusion in the calculations of this
paper.  \citet{xrb} also discussed a volume discrepancy factor that is
designed to drive the flow back onto the constraint dictated by the
equation of state.  For the present simulations we run with this off
(equivalent to $f = 0$ in Eq.~13 of \citealt{xrb}).

\begin{figure}
\centering
\includegraphics[height=8.0in]{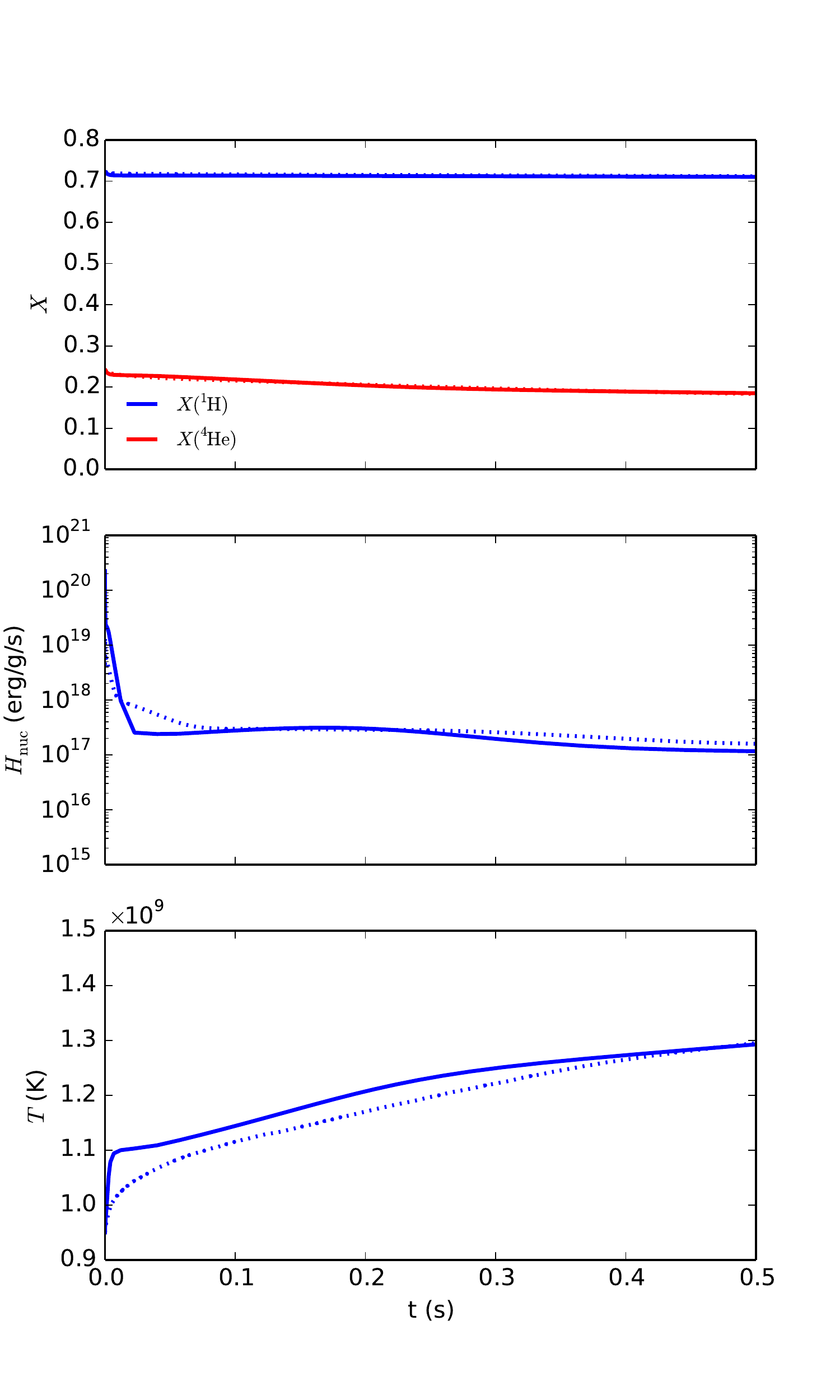}
\caption{\label{fig:rpcomp_low} Comparison of the thermodynamic
  evolution for a one-zone, isobaric, self-heating burn with the
  simple rp-process approximate network used for our simulations
  (solid lines) to a more extensive offline network (including nuclei
  up through Ge) from the Kepler code-base (dotted lines).}
\end{figure}

To assess the validity of the approximate network we perform a
one-zone, constant pressure, self-heating burn using the conditions
similar to those at the base of the accreted layer in an XRB: $T =
9.5\times10^8$ K, $\rho=5.93\times10^6$ g cm$^{-3}$, $X(\isotm{H}{1})
= 0.72$, $X(\isotm{He}{4}) = 0.24$, and the remaining abundance split
between \isot{O}{14} and \isot{O}{15} in a ratio comparable to their
respective $\beta^+$-decay times.  The solid lines in Figure
\ref{fig:rpcomp_low} show the early evolution of the hydrogen/helium
abundance (top), specific energy generation rate (middle), and
temperature (bottom) for our approximate network.  The dotted lines
show a comparison of a simple calculation with the same initial
conditions but using an adaptive reaction network from the Kepler
code-base, which included many nuclei up to Ge.  The two networks
agree quite well under these conditions, but the approximate network
initially tends to be $\lesssim10\%$ hotter than the larger Kepler
network.

\section{Initial Model and Simulation Parameters}\label{Sec:Initial Model}

We generate parametrized models of the accreted layer using
simulations run by the Kepler stellar evolution code as a guide.
Creating our own models alleviates the difficulty of reproducing the
proper convectively unstable regions in the atmosphere of the one-dimensional
model.  In particular, mapping from a Lagrangian grid to an Eulerian
grid, interpolating data to a constant mesh spacing, and slight
differences in the thermodynamics between codes can all lead to
differences in thermal/adiabatic gradients, and hence which region is
convectively unstable.  Additionally, spurrious features in the one-dimensional
model can cause difficulties while mapping onto an Eulerian grid and
maintaining HSE, especially in regions of sharp discontinuities in the
Lagrangian code.

Our simplified models and their generation are described in detail in
Appendix~\ref{app:models}.  Briefly, our models consist of a 1.4
$M_\odot$, 10 km, isothermal ($T=T_\mathrm{ns}$) ``neutron star''
substrate composed of heavy elements ($X = X_\mathrm{ns} \equiv$ pure
\isot{Ni}{56}) with a warm accreted layer of mixed fuel composition
($X=X_\mathrm{fuel}$), part of which is convectively unstable with an
isentropic gradient.  The transition between the neutron star and the
accreted layer is goverened by the density (\rhob) and temperature
(\Tb) at the base of the accreted layer.  The composition of the
accreted layer was approximated from an initial Kepler XRB model, but
is essentially a slightly metal-rich gas compared to solar
metallicity---to reflect prior burning---with the
\isot{O}{15}/\isot{O}{14} ratio set approximately by their respective
lifetime against $\beta^+$-decay.  Table~\ref{table:model_params}
gives the set of parameters that describe the models---refer to
Appendix~\ref{app:models} for the definitions.

Our first attempt at model creation used values of \rhob\ ($\sim
7\times10^5\ \gcc$) and \Tb\ ($\sim 7.5\times10^8$ K)  determined
from the original Kepler models.  When mapped into two dimensions,
these models exhibited weaker burning than their one-dimensional
counterparts.  This leads to a slowly increasing base temperature that
eventually peaked around $\lesssim 10^9$ K, and then remained roughly
constant.  As we found found in Paper~I with semi-analytic models, the
multidimensional convection generated in these models was much more
efficient at carrying heat than the mixing length theory prescription
used in the one dimensional Kepler models.  

\begin{figure}
\centering
\includegraphics[height=8.0in]{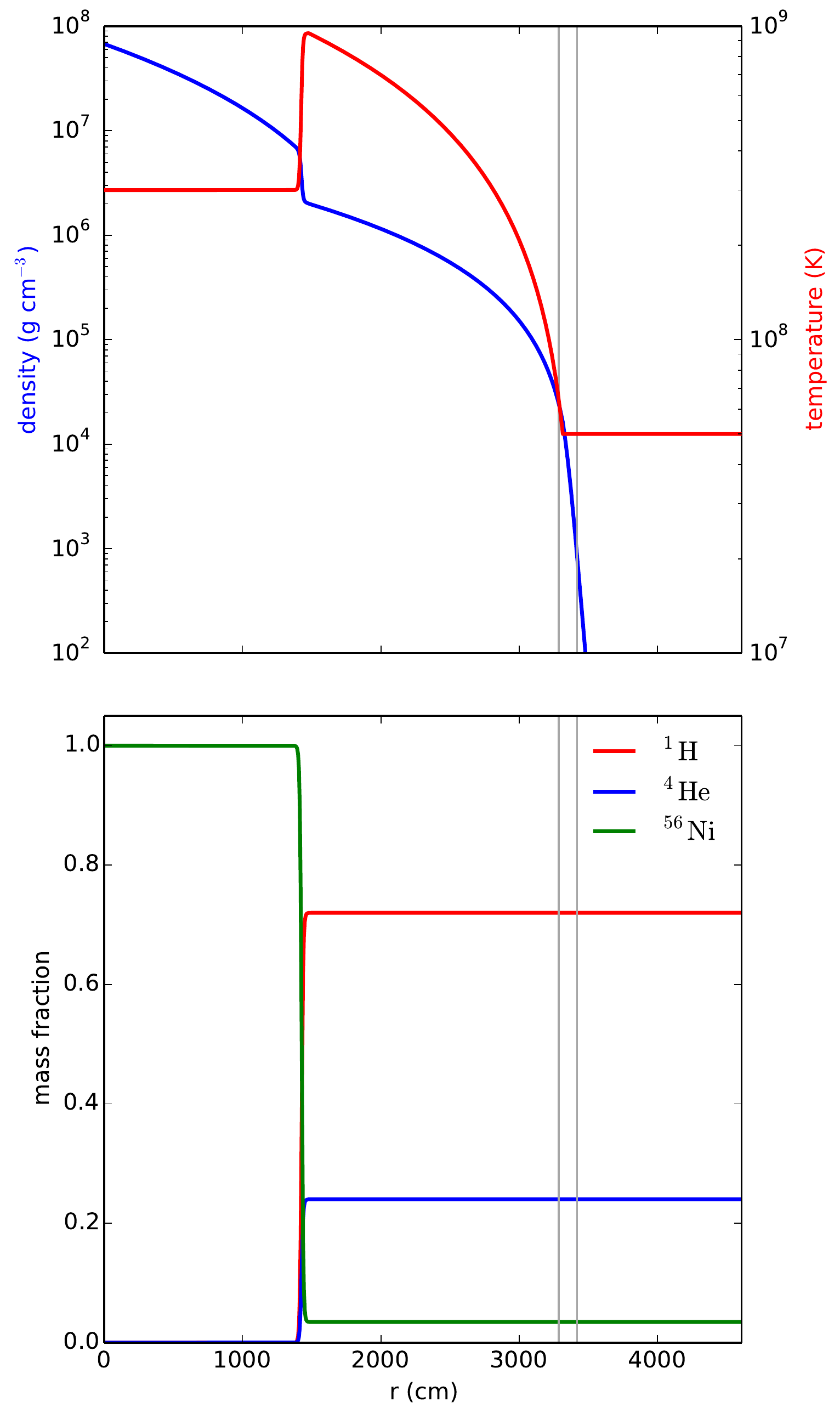}
\caption{\label{fig:initialmodeldens} Initial one-dimensional model
  for the simulations of this paper.  The vertical gray lines indicate
  the radial position of the start of the sponge
  ($r=r(\rho=f_\mathrm{sp}\rho_\mathrm{anel})$; leftmost line) and the
  cutoff density/anelastic cutoff ($r=r(\rho=\rho_\mathrm{anel})$;
  rightmost line).}
\end{figure}

To create more prolific burning, we therefore modified our initial
conditions to those given in Table \ref{table:model_params}.  Namely,
we artificially increased the density to $\rhob=2\times10^6~\gcc$ and
the temperature to $\Tb=9.5\times10^8$ K while keeping the composition
fixed.  Increasing the temperature and density has two main
consequences for the burning in the layer: 1) boosting the
triple-$\alpha$ reaction ($\propto\rho^2$), which is temperature
sensitive, relative to the $\beta$-limited HCNO cycle, and 2)
decreasing the branching ratios $\lambda_1$ and $\lambda_2$ (see
Figure \ref{fig:rprox_schematic}), which allows for an increase in
breakout from the HCNO cycle.  Ultimately, this quicker consumption of
\isot{He}{4} nuclei early on will slow the late-time evolution of the
burning as \ap\ reactions might not be frequent enough to bridge
$\beta^+$-decay waiting points.  However, our burning does not reach that
far into heavy elements so we are not concerned with this aspect here.
Figure \ref{fig:initialmodeldens} shows the initial density,
temperature and bulk composition profiles for the model used in our
calculations.

As with our simulations of convection in/on white dwarfs
\citep{subchandra,wdturb}, and also Paper~I,
we use a low density cutoff and a sponging
technique to control the behavior of the flow at the top of the atmosphere.
These are designed to suppress unwanted velocities in the
region above our accreted layer.  The low density cutoff acts to
change the behaviour of the low Mach number algorithm when the density
drops below a certain value.  In particular, when the density drops
below $\rho_\mathrm{anel}$, the divergence constraint on the velocity
field is altered to resemble that of the anelastic approximation,
$\nabla \cdot (\rho_0 U) = 0$.  There is another low density
cutoff, $\rho_\mathrm{cutoff}$, that is the density at which we fix
the ambient medium.  Both $\rho_\mathrm{anel}$ and
$\rho_\mathrm{cutoff}$ are runtime parameters, which we set to
$\rho_\mathrm{cutoff} = \rho_\mathrm{anel} = 10^3~\gcc$ in the present
simulations.  The location where the density first drops below the
cutoffs is shown as the right-most thin vertical gray line in Figure
\ref{fig:initialmodeldens}.

For the sponge, we pick a multiple, $f_\mathrm{sp}$, of the anelastic
cutoff to define the density at which the sponge turns on,
$\rho_\mathrm{sp} = f_\mathrm{sp} \rho_\mathrm{anel}$.  The sponge is
designed to be in full force when the density drops to
$\rho_\mathrm{anel}$.  Here we use a simplified version of the
sponging as compared to our previous work:
\begin{equation}
s = \left \{ \begin{array}{ll}
   1,  & \mathrm{if~} r < r(\rho_\mathrm{sp}) \\
   \frac{1}{2} (1 - s_\mathrm{min} ) \cos \left ( 
     \pi \frac{r - r(\rho_\mathrm{sp})}
          {r(\rho_\mathrm{anel}) - r(\rho_\mathrm{sp})} \right )
   + \frac{1}{2} ( 1 + s_\mathrm{min}), & \mathrm{if~} r(\rho_\mathrm{sp}) \le r < r(\rho_\mathrm{anel}) \\
   s_\mathrm{min}, & \mathrm{otherwise,}
\end{array} \right .
\end{equation}
and after the velocity is advanced, the sponge is applied as:
\begin{equation}
U^{n+1} \rightarrow s U^{n+1}.
\end{equation}
For all the simulations presented here, we choose $f_\mathrm{sp} =
25$, and $s_\mathrm{min} = 0.01$.  The left-most vertical gray line in
Figure \ref{fig:initialmodeldens} marks the location of the sponge
start, $r(\rho_\mathrm{sp})$ (i.e.\ where $\rho =
f_\mathrm{sp}\rho_\mathrm{anel}$).

To seed the convection, we add a number, $n_\mathrm{vort}$, of small
vortices to the initial conditions at a fixed height,
$r_\mathrm{vort}$, near the base of the convective region.  The
vortices are spaced equally across the domain, and the
orientation---clockwise vs.\ counterclockwise---is altered every other
vortex.  Each vortex is Gaussian in form with the velocity
perturbations given as
\begin{eqnarray*}
  \delta u_i(x,r) &=& -\left(-1\right)^iA_\mathrm{vort}(r - r_\mathrm{vort})\exp\left(-\frac{d_i^2}{\sigma^2}\right) \\
  \delta v_i(x,r) &=& \left(-1\right)^iA_\mathrm{vort}(x - x_{\mathrm{vort},i})\exp\left(-\frac{d_i^2}{\sigma^2}\right),
\end{eqnarray*}
where $u$ and $v$ are the horizontal and vertical components of the
velocity, respectively, $A_\mathrm{vort}$ is the amplitude of the
perturbation, $(x_{\mathrm{vort},i},r_\mathrm{vort})$ is the center of
the $i$th vortex, $d_i = \left[(x-x_{\mathrm{vort},i})^2 +
  (r-r_\mathrm{vort})^2\right]^{1/2}$ is the distance from the center
of $i$th vortex, and $\sigma$ is the size of the vortex.  The
superposition of all vortices is used to determine the initial
velocity field; that is
\begin{equation}\label{eq:velpert}
(u,v) =\left(\sum\limits_{i=1}^{n_\mathrm{vort}}\delta
u_i,\sum\limits_{i=1}^{n_\mathrm{vort}}\delta v_i\right).
\end{equation}
For the simulations presented here we choose $r_\mathrm{vort} = 1475$
cm, $A_\mathrm{vort} = 1 $ km s$^{-1}$, $\sigma^2 = 200$ cm$^2$, and
$n_\mathrm{vort} = 16$ (each $x_{\mathrm{vort},i}$ is determined from
the domain width and $n_\mathrm{vort}$).  We note that the initial
convective region for this model spans roughly the region between 1470
cm $\lesssim r \lesssim$ 3350 cm.

\section{Results}\label{Sec:Results}

We perform simulations at three resolutions: 12, 6, and 3~cm
zone$^{-1}$.  The 12 and 6~cm runs are done with a unifom grid.  For
the 3~cm run, we added a single level of refinement to the 6~cm model
in a region encompassing the convective zone.  We do not 
dynamically change the grid with time; this allows us to
optimize the load balancing of the simulation.  Each of these
simulations used a two-dimensional grid of size 1536 cm $\times$ 4608
cm.  A double-wide model at 6 cm zone$^{-1}$ resolution was also
created and mapped onto a grid of 3072 cm $\times$ 4608 cm.  This
fourth model comprises the bulk of the study in this paper as it was
evolved the furthest in time.

All simulations here assumed plane-parallel geometry with a constant
gravitational acceleration of $g = -2.45\times10^{14}$ cm s$^{-2}$.
Periodic boundary conditions were used in the lateral, $x$-direction.
The vertical, or $r$-direction, upper boundary was an outflow boundary
while the lower boundary used slip-wall conditions.

\subsection{General Trends and Resolution Dependence}
\label{sec:trends}

As in our simulations of Paper~I, a transient event occurs at the
onset of convective overturn.  This transient causes relatively large
velocity spikes until the system adjusts to a more steady-state
convective flow field.  Based on our experience in previous studies of
convection with Maestro \citep{wdturb,subchandra}, including an
initial velocity perturbation, such as that in Equation
\ref{eq:velpert}, helps to minimize the transient by giving the system
a small ``kick'' allowing the energy to be more quickly dispersed via
advection.  This is different from the approach of Paper~I where we
let solver noise seed convection.  After a few turnover times
($\lesssim 10^{-3}$ s), however, the convective pattern does not
remember how it was initiated, and therefore the results presented
here are not sensitive to the choice of initial perturbations.

\begin{figure}
  \centering
  \includegraphics[width=\textwidth]{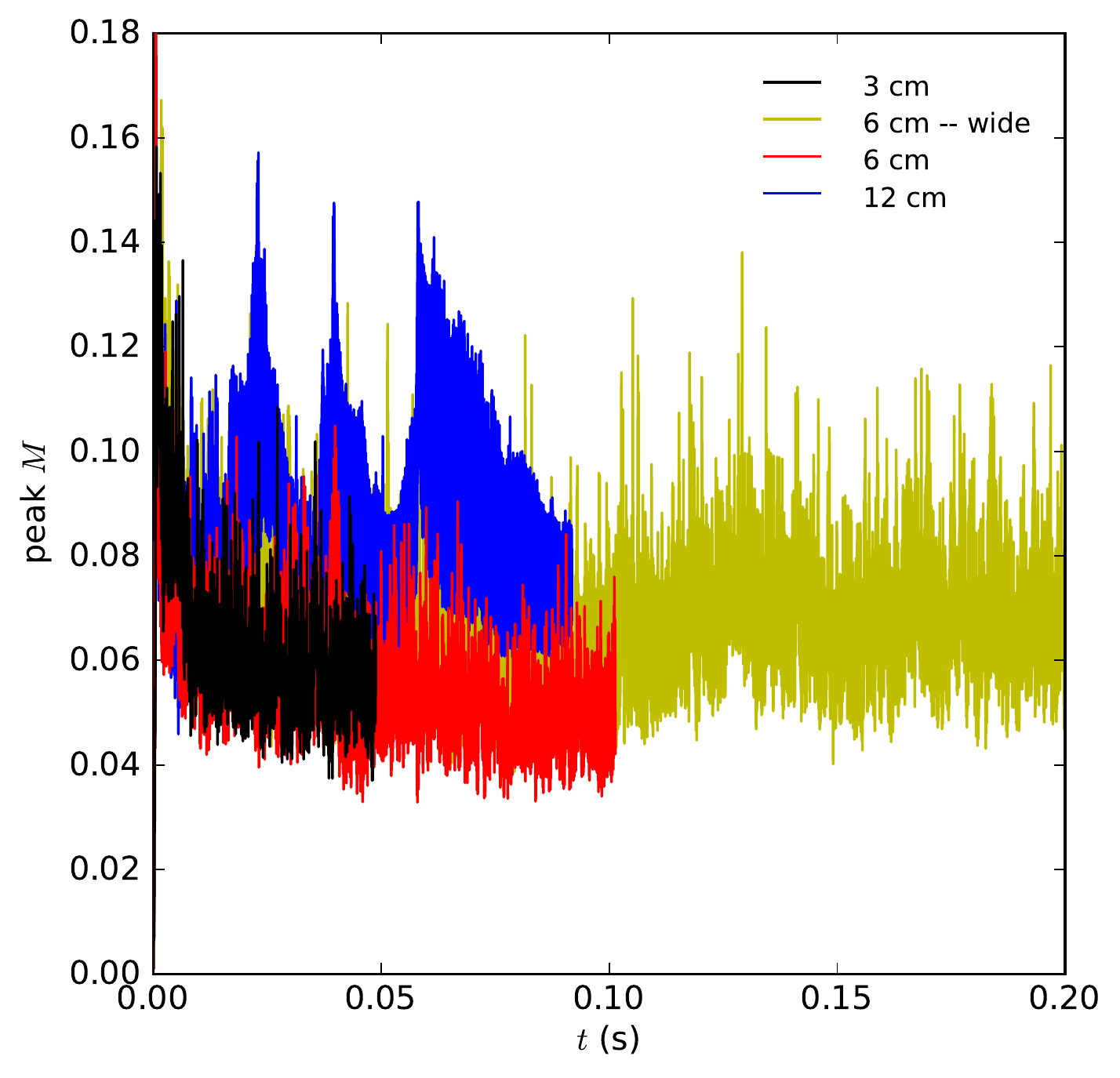}
  \caption{\label{fig:resstudyM} Resolution study showing the maximum
    Mach number on the grid.  There is good agreement between the 3
    and 6~cm zone$^{-1}$ models, suggesting that we are nearly
    converged.  We also see that peak Mach number systematically
    decreases with increasing resolution, likely due to better
    capturing of the peak of burning at the base of the accreted
    layer.}
\end{figure}

Figure \ref{fig:resstudyM} shows the peak Mach number, $M$, in the
domain as a function of time for all four of our models.  At very
early time, the Mach number peaks to about 0.18 for all simulations
because of the transient discussed above.  By about $t=0.01$ s, the
system has relaxed to a steady-state convective flow.  While the 12~cm
zone$^{-1}$ run differs markedly from the others, the 3 and 6~cm
zone$^{-1}$ runs appear to be converged.  Also note that the Mach
number is generally smaller for the higher resolution runs.  We
suspect that the primary source of the differences with resolution is
in the reaction source term.  The energy generation is strongly peaked
near the base of the burning layer, and, as a result, a coarse
simulation will underestimate the total energy generation (since the
coarse-zone cell-center is at a lower temperature than the a fine-zone
cell-center at the base of the layer).  This was also discussed in
Paper~I in the context of pure triple-$\alpha$ burning, which required
a very fine spatial resolution due to its high
temperature-sensitivity.  Also similar to the results of Paper~I, the
time-averaged peak velocity remains less than 10\% of the sound speed,
validating the use of a low Mach number approximation method for XRBs.

\begin{figure}
  \centering
  \includegraphics[width=\textwidth]{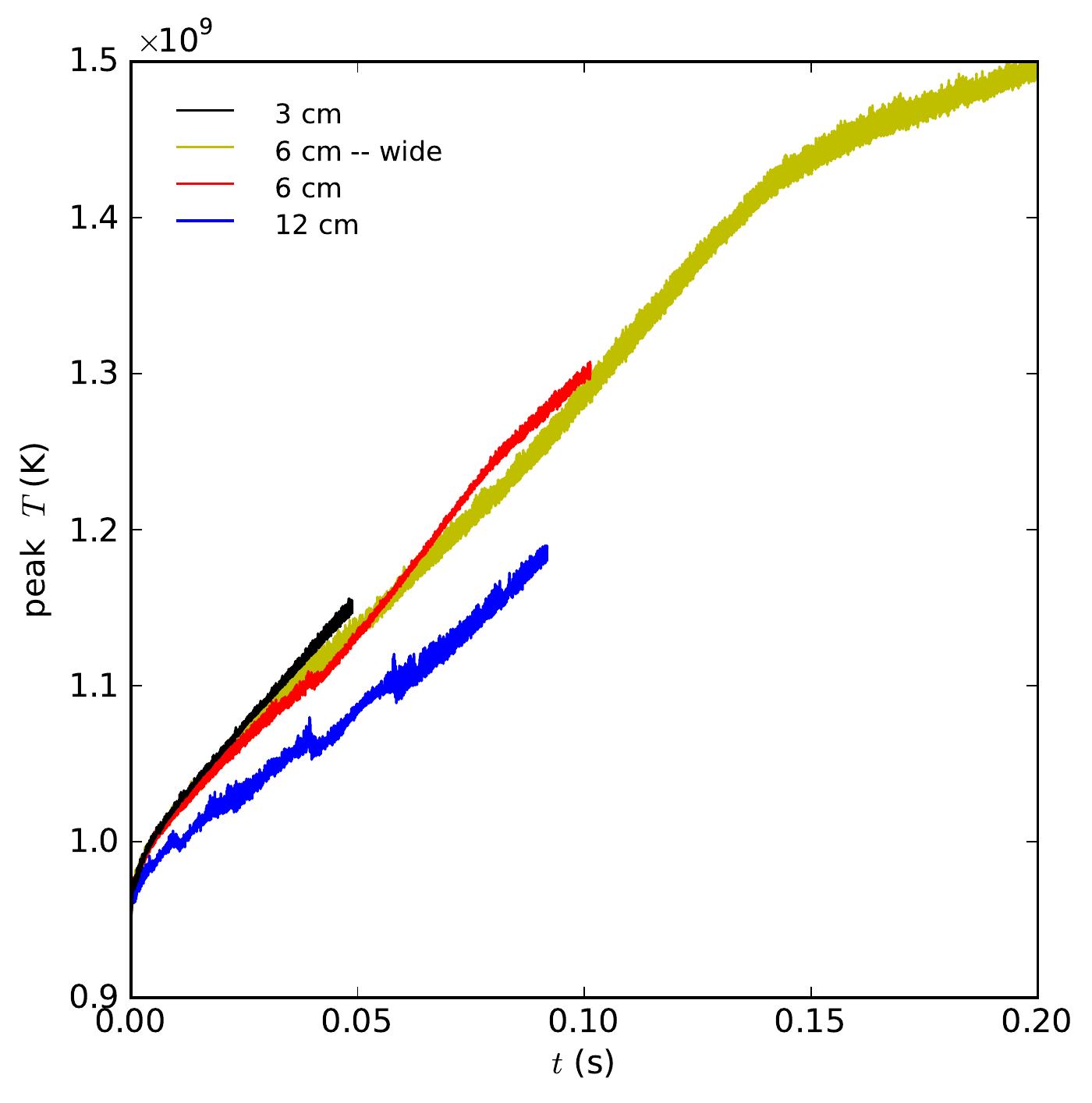}
  \caption{\label{fig:resstudy} Resolution study showing the peak
    temperature in the domain as a function of time.  Again, here we
    see reasonable agreement in the trend of peak temperature
    vs.\ time between the 3 and 6~cm zone$^{-1}$ models, suggesting
    that we are nearly converged.  The coarse 12~cm zone$^{-1}$ run
    tends to be cooler due to the increased amount of overshoot and
    poorly-captured burning profile (see text).}
\end{figure}

Similarly, Figure \ref{fig:resstudy} shows the peak temperature in the
domain as a function of time for all four runs.  Again we
see---excepting the 12~cm zone$^{-1}$ run---the temperature evolution
of the models is quite similar, agreeing to within $\lesssim 5\%$.  In
addition to underestimating the energy generation, the coarse model
with 12~cm zone$^{-1}$ resolution also tends to overestimate the
amount of convective overshoot.  This can also be inferred from Figure
\ref{fig:resstudyM}: the overall larger velocities allow for parcels
of fluid on ballistic trajectories to more readily penetrate the
convective boundary thereby increasing the heat flux {\it away} from
the base of the accreted layer, and increasing the rate at which it
can cool.  

Of the three different resolutions we tested, the 6~cm zone$^{-1}$
simulations offer the best tradeoff between accuracy and computational
cost.  The narrow-domain 6~cm zone$^{-1}$ model, however, has a domain
width comparable to the initial vertical extent of the convective
region.  This forces the convective rolls to assume an aspect ratio
$\lesssim 1$, which alters the dynamics and allows for the
$\lesssim30\%$ difference in the time-averaged $M$ between the normal
6~cm zone$^{-1}$ model and the wide model.  The double-wide domain
alleviates the aspect ratio forcing for the duration of the
simulations presented here.  We therefore focus on the 6~cm
zone$^{-1}$, wide-domain simulation for the remainder of this paper.

\subsection{Convection in the Wide Domain}

\subsubsection{Early Adjustment}
Figure \ref{fig:initconv} shows the transient adjustment phase
discussed in Section \ref{sec:trends} as seen in three different
variables: magnitude of vorticity (left column), Mach number (middle
column), and specific energy generation rate (right column).  The
initial conditions are in the top row; note that the vortices of
Equation \ref{eq:velpert} are just barely discernible on the vorticity
color scale.  After $3\times10^{-4}$ s (middle row), the nonlinear
interactions of the perturbations give rise to a handful of dominant
plumes that provide the large Mach number of the transient startup.
Another $3\times10^{-4}$ s later ($t=6\times10^{-4}$ s; bottom row)
and the convective pattern has fully filled the region that was
originally Schwarzschild-unstable in the initial model.  We see
localized vortices, and where the flow converges near the bottom of
the convective layer, we also have a localized increased burning rate.

\subsubsection{Well-defined Convection Region}
\begin{figure}
  \centering
  \includegraphics[width=\textwidth,clip,trim = 0 32mm 0  0  ]{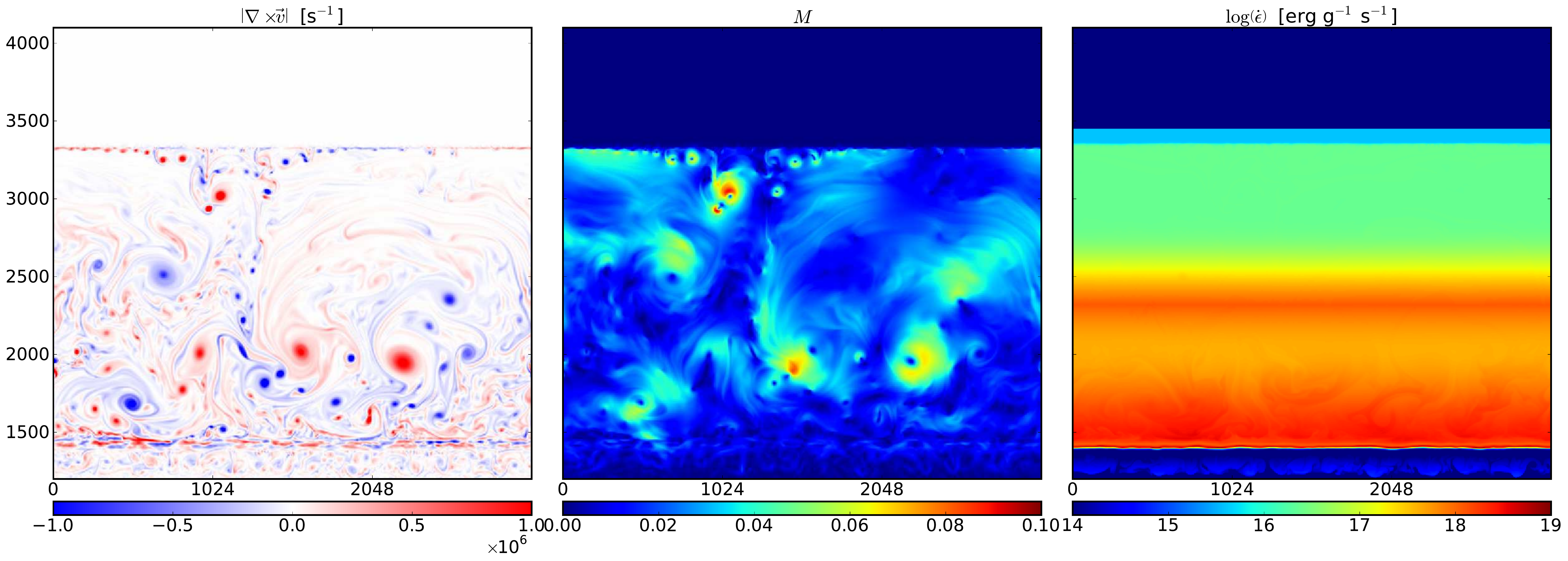}
  \includegraphics[width=\textwidth,clip,trim = 0 32mm 0 11mm]{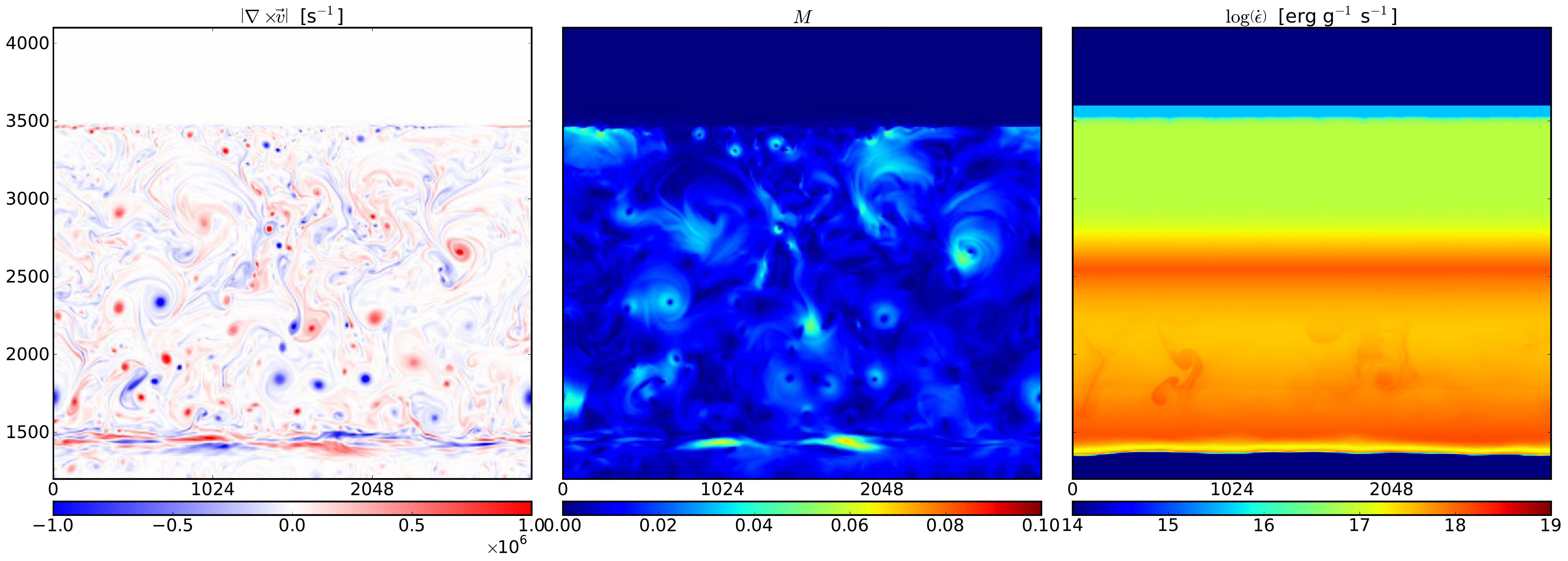}
  \includegraphics[width=\textwidth,clip,trim = 0 32mm 0 11mm]{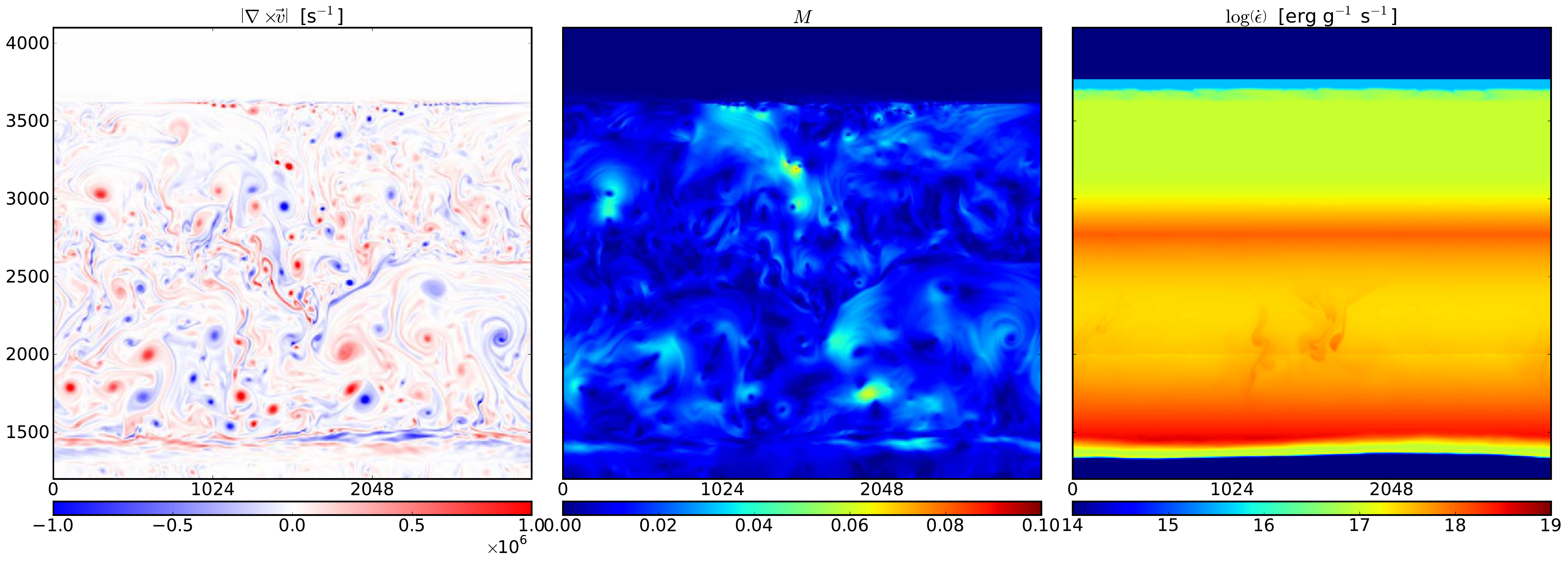}
  \includegraphics[width=\textwidth,clip,trim = 0  0   0 11mm]{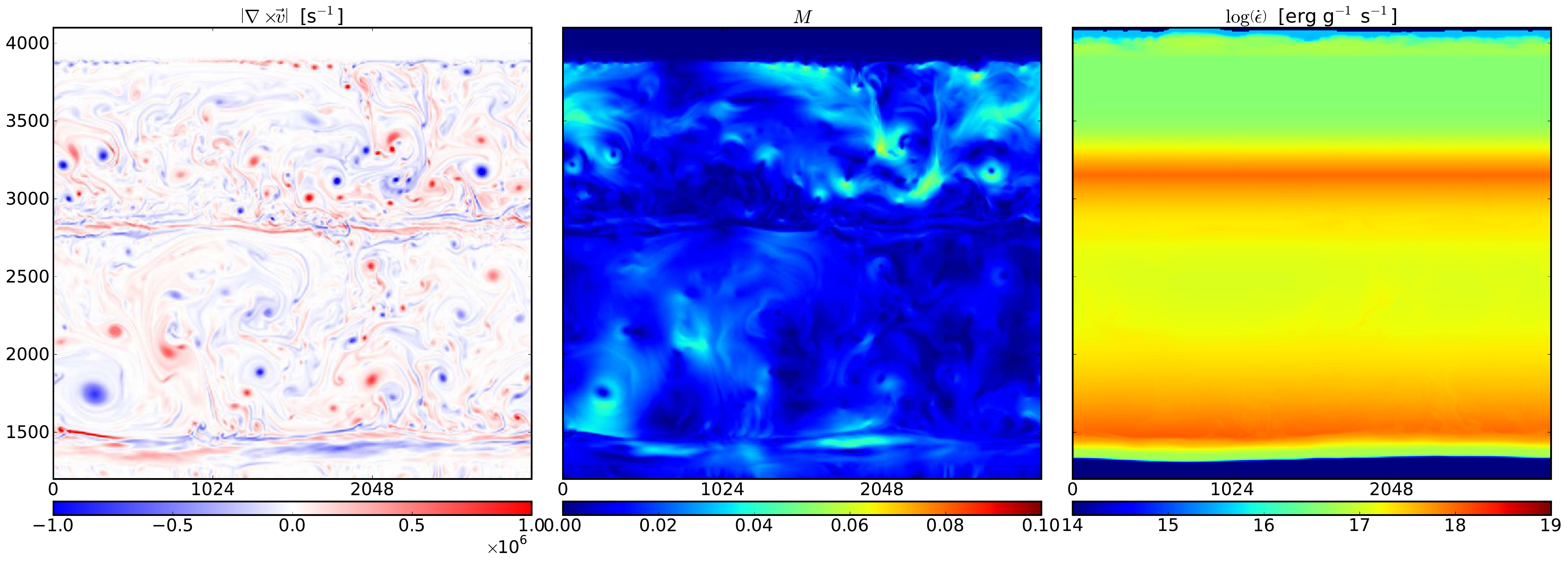}
  \caption{\label{fig:evolconv} Convective evolution of magnitude of
    vorticity (left), Mach number (middle), and logarithm of the
    specific energy generation rate (right) at $t=5\times10^{-3}$~s,
    $5\times10^{-2}$~s, $10^{-1}$~s, and $2\times10^{-1}$~s from top
    to bottom.  The secondary peak in energy generation rate tends to
    dominate the burning after $\sim10^{-1}$~s of evolution, and the
    convective region splits into layers shortly thereafter.}
\end{figure}

Figure \ref{fig:evolconv} shows the same evolution as Figure
\ref{fig:initconv}, except at later times ($5\times10^{-3}$,
$5\times10^{-2}$, $10^{-1}$, and $2\times10^{-1}$~s, from top to
bottom).  As is well known in two-dimensional simulations, smaller
vortices tend to merge into larger vortices.  Some of the smaller
scale features are smoothed out and the burning becomes (laterally)
very uniform within the burning region.  It is also clear from Figure
\ref{fig:evolconv} that the convective region expands both radially
inward and outward, with the extent of the former being much less than
that of the of latter.

\begin{figure}
  \centering
  \includegraphics[width=\textwidth]{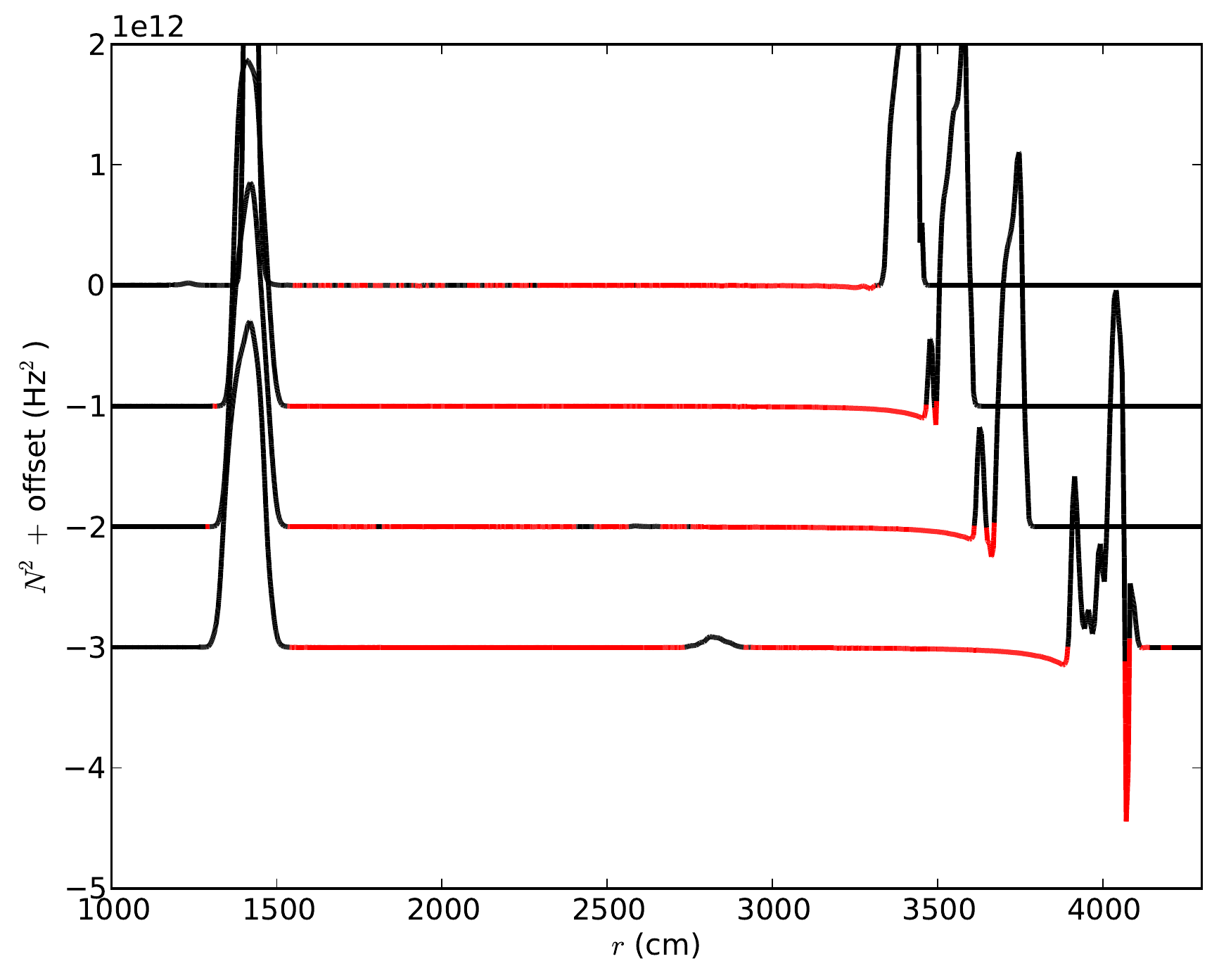}
  \caption{\label{fig:conv} Square of the Brunt-V\"ais\"al\"a
    frequency, $N^2$, profiles {\em as calculated from laterally
      averaged quantities}.  Each profile is offset by
    $10^{12}$~Hz$^2$ for clarity; the times for each profile are
    $t=5\times10^{-3},\ 5\times10^{-2},\ 10^{-1},$ and
    $2\times10^{-1}$ s from top to bottom.  Regions of convecitve
    instability ($N^2 <0$) have been highlighted red.  The layer
    formation in the convective region is prominent in the latest
    profile as the small island of stability (positive $N^2$) around
    $r=2800$~cm.}
\end{figure}

The expansion of the convectively unstable region can also been seen
by looking at the Brunt-V\"ais\"al\"a or buoyancy frequency, $N$, for a
stratified atmosphere (see, for example, Chapter 6 in
\citet{Kipp_Weigert}):
\begin{equation} \label{eq:bvf}
  N^2 = -g \left(\frac{d\ln\rho}{dr} -
  \left.\frac{d\ln\rho}{dr}\right|_{\mathrm{ad}}\right),
\end{equation}
where the subscript `ad' means along an adiabat, and we are assuming
instability to Schwarzschild convection. Equation \ref{eq:bvf} gives
the square of the frequency at which a fluid element will oscillate if
displaced from its equilibrium position ($\propto e^{iNt}$).  When
$N^2 < 0$, then the buoyancy frequency is purely imaginary, giving
rise to an instability.  Figure \ref{fig:conv} shows profiles of the
square of the Brunt-V\"ais\"al\"a frequency {\em as calculated from
  the laterally averaged quantities}---i.e.\ with the gradients in
Equation \ref{eq:bvf} being calculated as 
\begin{displaymath}
  \frac{d\ln\rho}{dr}\longrightarrow \frac{d\ln\langle\rho\rangle}{dr}
\end{displaymath}
and
\begin{displaymath}
  \left.\frac{d\ln\rho}{dr}\right|_{\mathrm{ad}} \longrightarrow
  \left.\frac{d\ln\rho(\langle p\rangle,\langle T\rangle,\langle
    X\rangle)}{dr}\right|_{\mathrm{ad}},
\end{displaymath}
where angle brackets denote laterally averaged quantities.  The
different curves are, from top to bottom, profiles at $t =
5\times10^{-3},\ 5\times10^{-2},\ 10^{-1},\ \mathrm{and\ }
2\times10^{-1}$ s, each being offset for clarity.  The regions of
instability ($N^2 < 0$) for each curve have been highlighted red.  

In an average sense, the upper boundary of the convective region
pushes radially outward, consistent with the colormap plots of Figure
\ref{fig:evolconv}.  However, at any given instance, the tenuous material
just outside the upper convective boundary gives rise to fluctuations
in entropy (or $N^2$) that make a sharp distinction of the boundary
difficult.  Indeed, the plots of $N^2$ at the upper boundary have much
more noise than at the lower boundary, where there exists a sharp
transition between neutron star and atmosphere.  The lower boundary is
always marked by a large (positive) $N^2$.  This results in stable
displacements of fluid elements that give rise to internal gravity
waves such as the wave-like features at the base of the
convective region in the plots of energy generation in Figure
\ref{fig:evolconv}.  

A peculiar feature evident in Figure \ref{fig:conv} at late time
(bottom curve) is a small region of convective stability around $r =
2850$ cm that splits the single convective zone into two layers of
convection.  This can be seen in either the vorticity or Mach number
plots of Figure \ref{fig:evolconv} at the same time (last row), which
shows the presence of two layers.  Indeed, a close examination of the
same plots at at an even earlier time, $t=10^{-1}$ s (third row of
Figure \ref{fig:evolconv}), hints at layer formation.

\subsubsection{Burning and Breakout}
\begin{figure}
  \centering
  \includegraphics[width=\textwidth,clip,trim = 0 32mm 0  0  ]{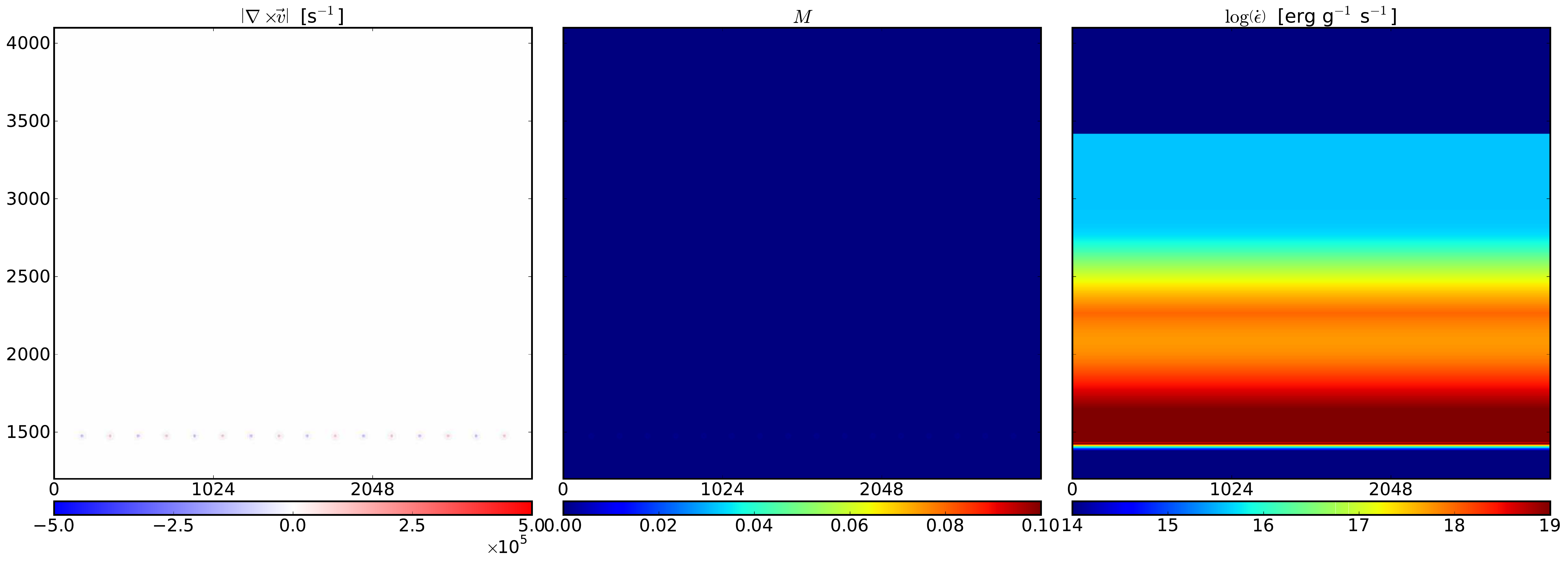}
  \includegraphics[width=\textwidth,clip,trim = 0 32mm 0 11mm]{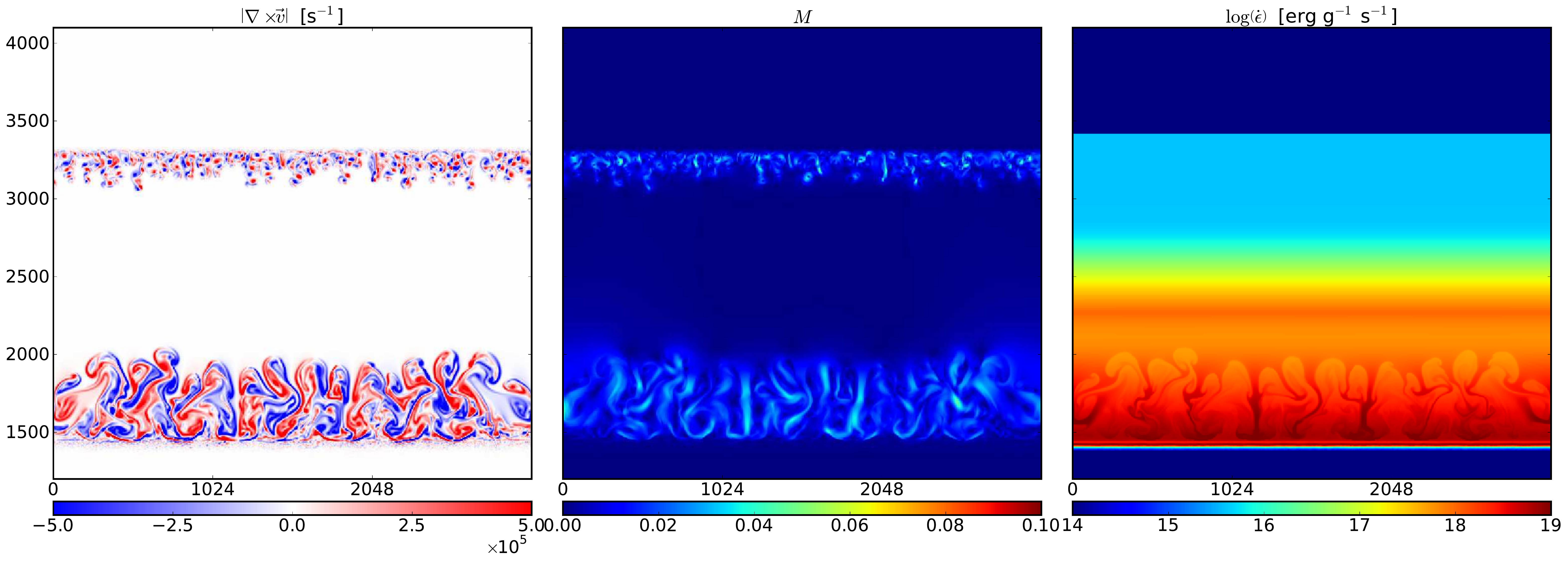}
  \includegraphics[width=\textwidth,clip,trim = 0  0   0 11mm]{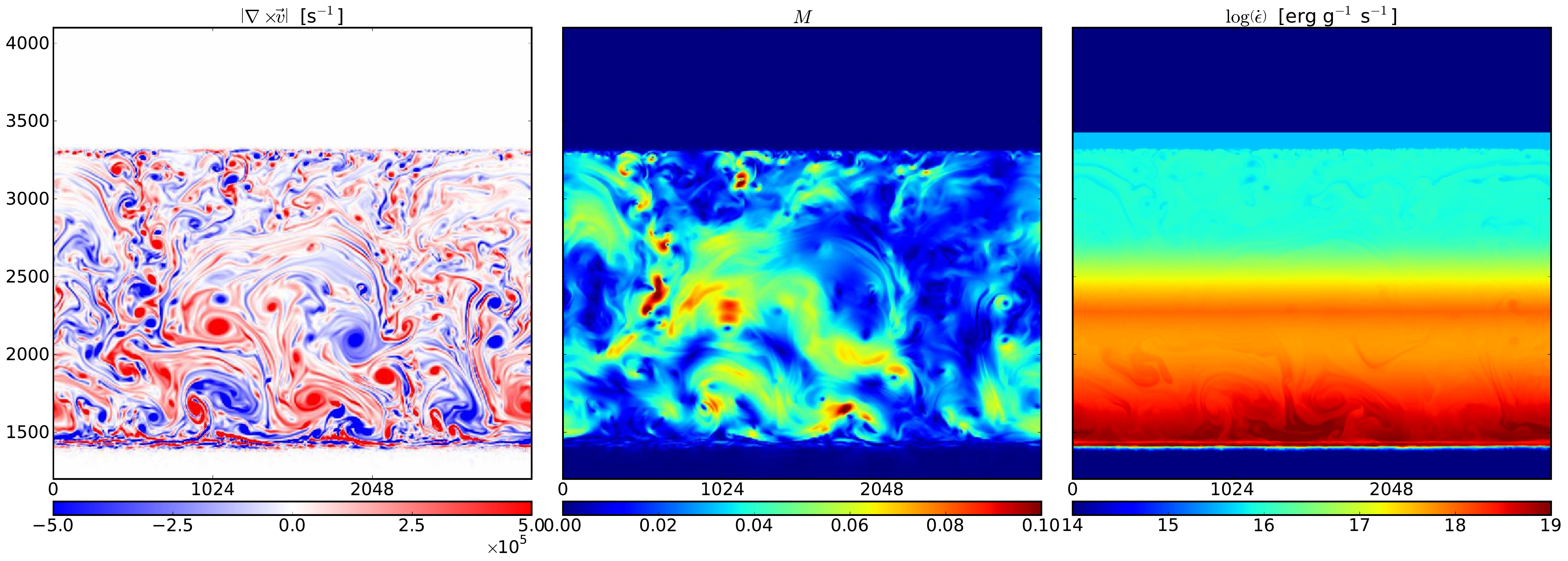}
  \caption{\label{fig:initconv} Early adjustment of the system shown
    in magnitude of vorticity (left), Mach number (middle), and
    logarithm of the specific energy generation rate (right) at
    $t=0$~s, $3\times10^{-4}$~s, and $6\times10^{-4}$~s from top to
    bottom.  By $t\sim5\times10^{-4}$~s the flow pattern has filled
    the entire convective region.  Note the presence of a secondary
    local maximum in energy generation rate in the initial conditions
    around $r\sim2400$~cm.}
\end{figure}

\begin{figure}
  \centering
  \includegraphics[width=\textwidth]{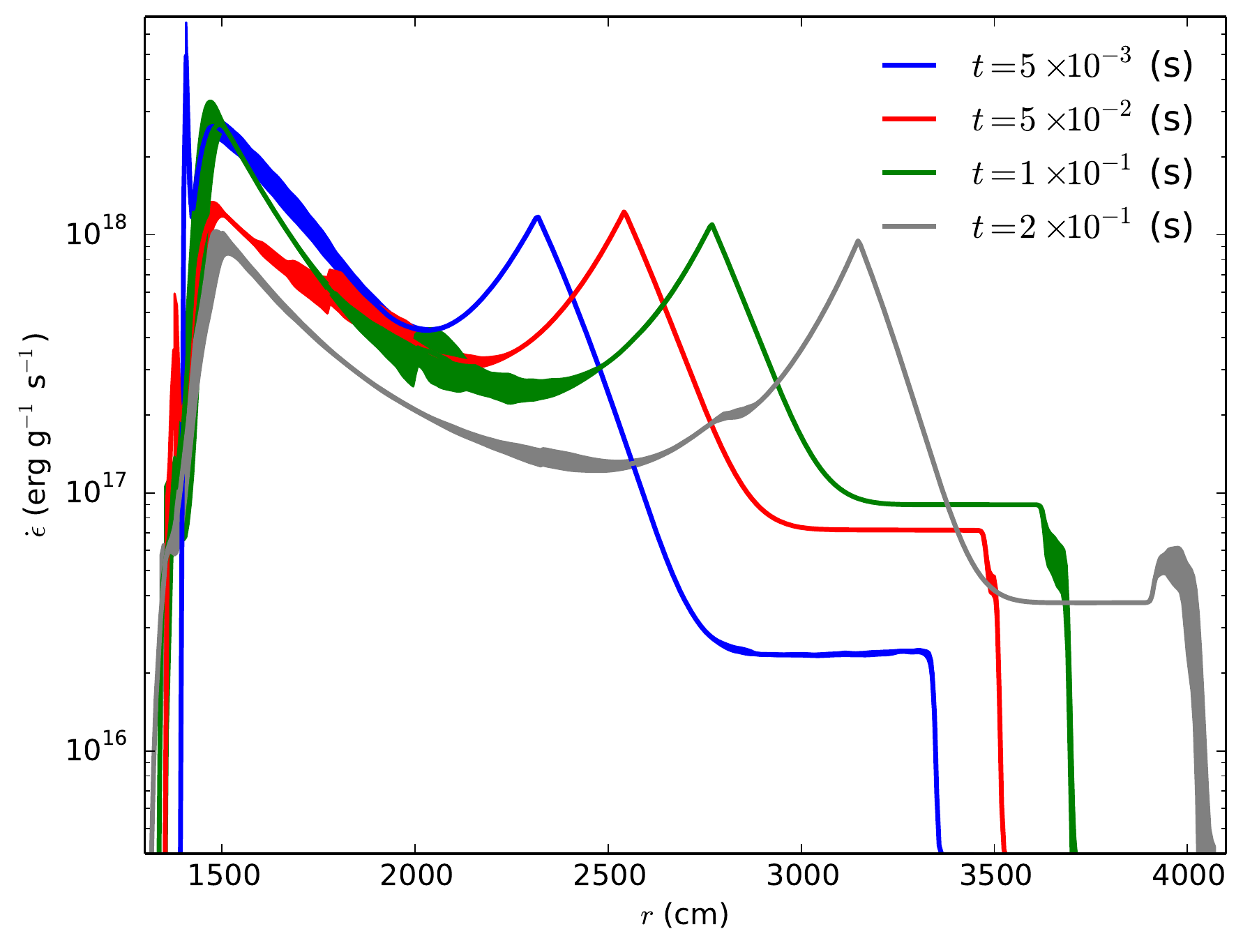}
  \caption{\label{fig:enuclate} Specific energy generation rate,
    $\dot\epsilon$, profiles at late time.  Each profile shows the
    lateral average at a given radius plus or minus the
    root-mean-square deviation from the lateral average.  Note that as
    time progresses, the second peak in $\dot\epsilon$ becomes
    relatively more prominent and moves further out in the
    atmosphere.}
\end{figure}

The location of layer splitting in Figure \ref{fig:evolconv} is
coincident with, but below a local maximum in the energy generation
rate.  Looking back at the initial conditions for the energy
generation rate in the top row of Figure \ref{fig:initconv}, we see
that indeed there is a secondary, weaker peak above the base of the
atmosphere.  This peak tends to strengthen with time while the peak of
burning at the base of the accreted layer weakens (as seen through
Figures \ref{fig:initconv} and \ref{fig:evolconv}).  Figure
\ref{fig:enuclate} shows the lateral average ($\pm 1\sigma$) of the
energy generation rate at the same time as the snapshots in Figure
\ref{fig:evolconv}.  By $t = 10^{-1}$ s, the two peaks have comparable
energy output, while at later times what was initially the secondary
peak slightly dominates the burning at the base of the accreted layer.
The secondary peak also appears to move radially outward in the
atmosphere at about the same rate as the expansion of the convective
zone $\sim 60$ m s$^{-1}$.

\begin{figure}
  \centering
  \includegraphics{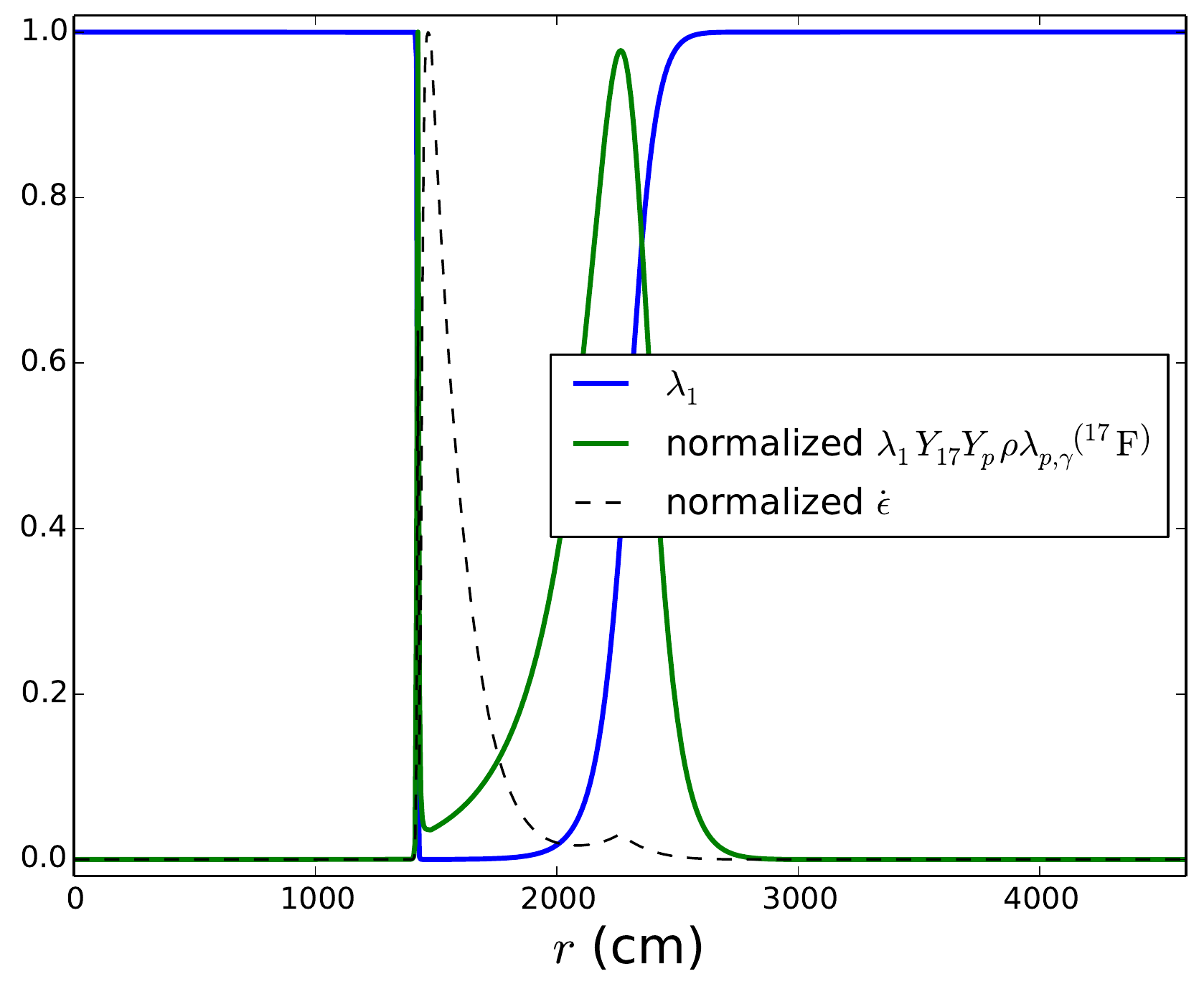}
  \caption{\label{fig:br} Profiles of a dominant rp-process breakout
    branching ratio, $\lambda_1$ (blue; see text and/or Figure
    \ref{fig:rprox_schematic} for definition)), normalized energy
    generation rate (dashed), and normalized rate at which
    \isot{F}{17} (plus a proton) is converted directly to \isot{O}{15}
    (plus an $\alpha$) (red) in the approximate network.  This
    approximate reaction chain is responsible for the secondary peak
    in energy generation rate seen in Figures \ref{fig:initconv},
    \ref{fig:evolconv}, and \ref{fig:enuclate}, and the split of the
    convective region into two layers.}
\end{figure}

The presence of the secondary peak is a feature strenghtened by the
approximations in the network used here.  In general, in the accreted
layer the density and temperature are {\em decreasing} functions of
radius.  How, then, can a constant composition profile produce a
non-monotonic energy generation rate profile?  The answer lies in the
branching ratio between rp-process breakout and returning to the HCNO
cycle in the approximate network.  In particular, for the initial
conditions, Figure \ref{fig:br} shows in blue the branching ratio,
$\lambda_1$, which was defined in Figure \ref{fig:rprox_schematic} as
the fraction of \isot{Ne}{18} that $\beta^+$-decays to \isot{F}{18}
faster than it can $\alpha$-capture.  The $\beta^+$-decay rate
is---for lack of better physics of $T$-dependent weak decays---a
constant, whereas the $\alpha$-capture has to overcome a Coulomb
barrier.  This causes $\lambda_1$ to increase with decreasing
temperature and we see a sharp transition in the branching ratio
around $r\approx2200$ cm.  Note that the branching ratio is unity in
the neutron star substrate as there is no \isot{He}{4} available for
capture on \isot{Ne}{18}.  Also shown in Figure \ref{fig:br} in red is
the normalized rate at which \isot{F}{17} (plus a proton) is converted
{\em directly} to \isot{O}{15} (plus an $\alpha$) in the approximate
network---this rate is goverened by the \isot{F}{17} can
proton-capture to \isot{Ne}{18}
($\lambda_{p,\gamma}\left(\isotm{F}{17}\right)$) and the fraction of
\isot{Ne}{18} that $\beta^+$-decay ($\lambda_1$).  This particular
rate coincides precisely with the secondary peak in the energy
generation rate, which is shown (normalized) as the dashed line.  It
is this approximate reaction sequence that is responsible for the
secondary peak in energy generation rate seen in Figures
\ref{fig:initconv},\ref{fig:evolconv}, and \ref{fig:enuclate}, and for
the formation of the two layered convective structure clearly present
at late time in Figure \ref{fig:conv}.  We stress that the formation
of distinct layers in the convective region is likely not physical,
but rather an artifact of the approximate network used in these
studies.

\begin{figure}
  \centering
  \includegraphics{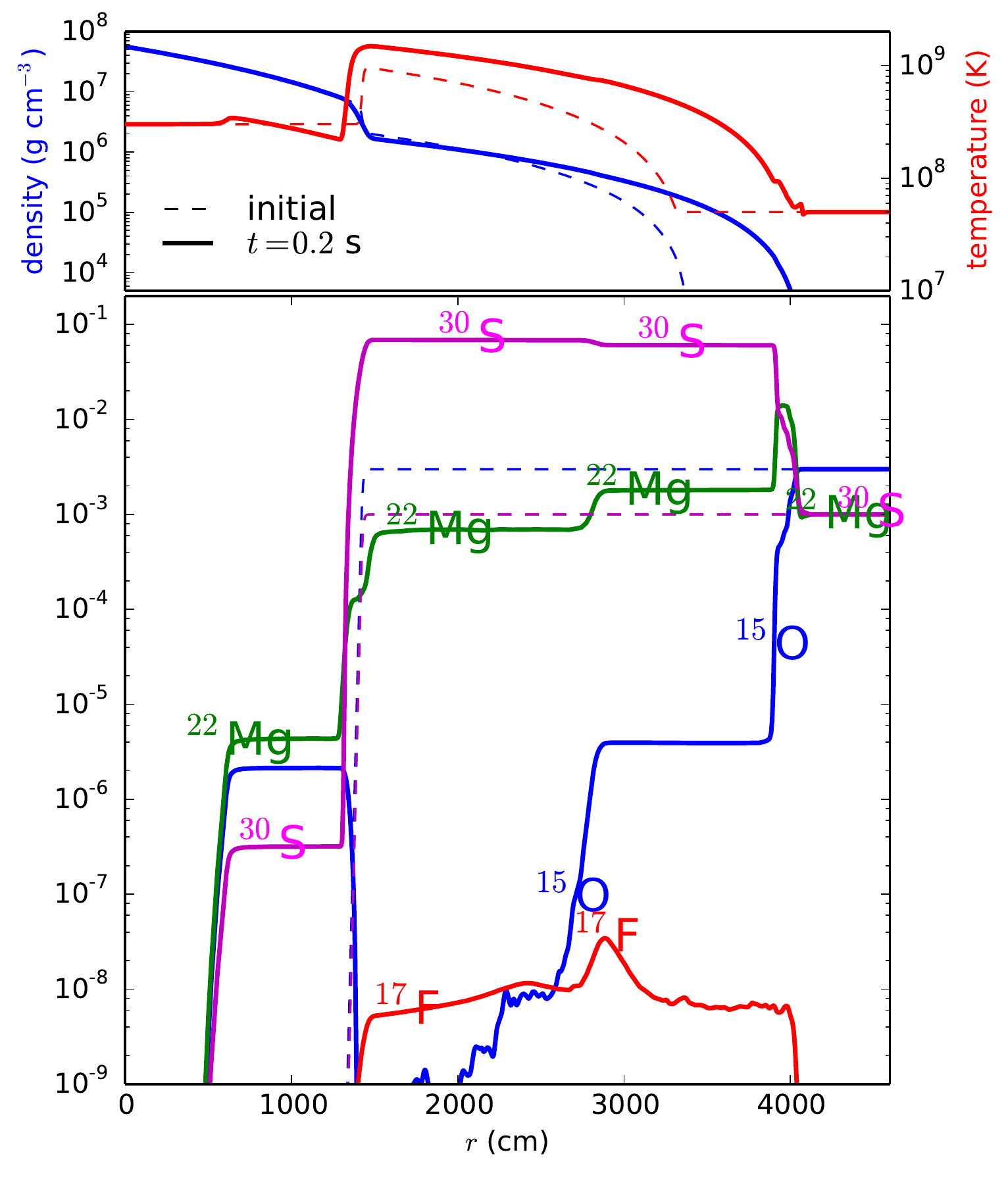}
  \caption{\label{fig:spec} Density, temperature, and various species
    profiles both at the start (dashed lines) and end ($t=0.2$~s;
    solid lines) of the simulation for the wide 6 cm zone$^{-1}$
    model.  The density and temperature profiles show just how much
    the atmosphere has expanded during the burn.  The composition
    profiles show both regions of convective overshoot (relatively
    large abundances below $r\sim1500$ cm) as well as the split in
    convective layers around $r\sim2800$ cm.}
\end{figure}

Figure \ref{fig:spec} shows density, temperature, and various
composition profiles as a function of radius both at the beginning
(dashed lines) and end (solid lines) of the 6 cm zone$^{-1}$, wide
simulation.  Note that of the isotopes plotted, only \isot{O}{15} and
\isot{S}{30} had an initial abundance within the range of the plot
(see Table \ref{table:model_params}).  The density and temperature
profiles in the top panel show just how much the atmosphere has been
heated and puffed up in the short duration of the simulation.  The
outer layers of the atmosphere have been heated substantially from the
secondary energy generation peak.  Likewise, the composition profiles
betray the presence of layering in the atmosphere due to the secondary
burning peak.  The base of the accreted layer is around $r\sim 1500$
cm.  The relatively large abundance of, for example, \isot{O}{15},
\isot{Mg}{22}, and \isot{S}{30} below this region is due to convective
overshoot; similarly, although not shown, a signficant amount of
\isot{Ni}{56} is dredged up into the atmosphere from the neutron star
material via the overshoot.  In the first convective layer,
$1500\mathrm{\ cm}\lesssim r \lesssim 2800\mathrm{\ cm}$, the
\isot{Mg}{22} produced from prior burning is readily being converted
to \isot{S}{30}.  In the second layer, $2800 \mathrm{\ cm}\lesssim r
\lesssim 4000\mathrm{\ cm}$, where the branching ratio $\lambda_1$
favors \isot{O}{15} production from \isot{F}{17}, we see a relatively
larger abundance of \isot{O}{15}.  There is still a significant
abundance of \isot{Mg}{22} in this region, mainly due to mixing
throughout the convective region before it was split into two layers,
but also due to some burning of \isot{O}{14} (via $\lambda_2$) and
\isot{O}{16}.  The lower temperature and density in the second
convective layer means \isot{Mg}{22} is converted to \isot{S}{30} at a
slightly slower rate than in the in the first convective layer.

\begin{figure}
  \centering
  \includegraphics[width=\textwidth]{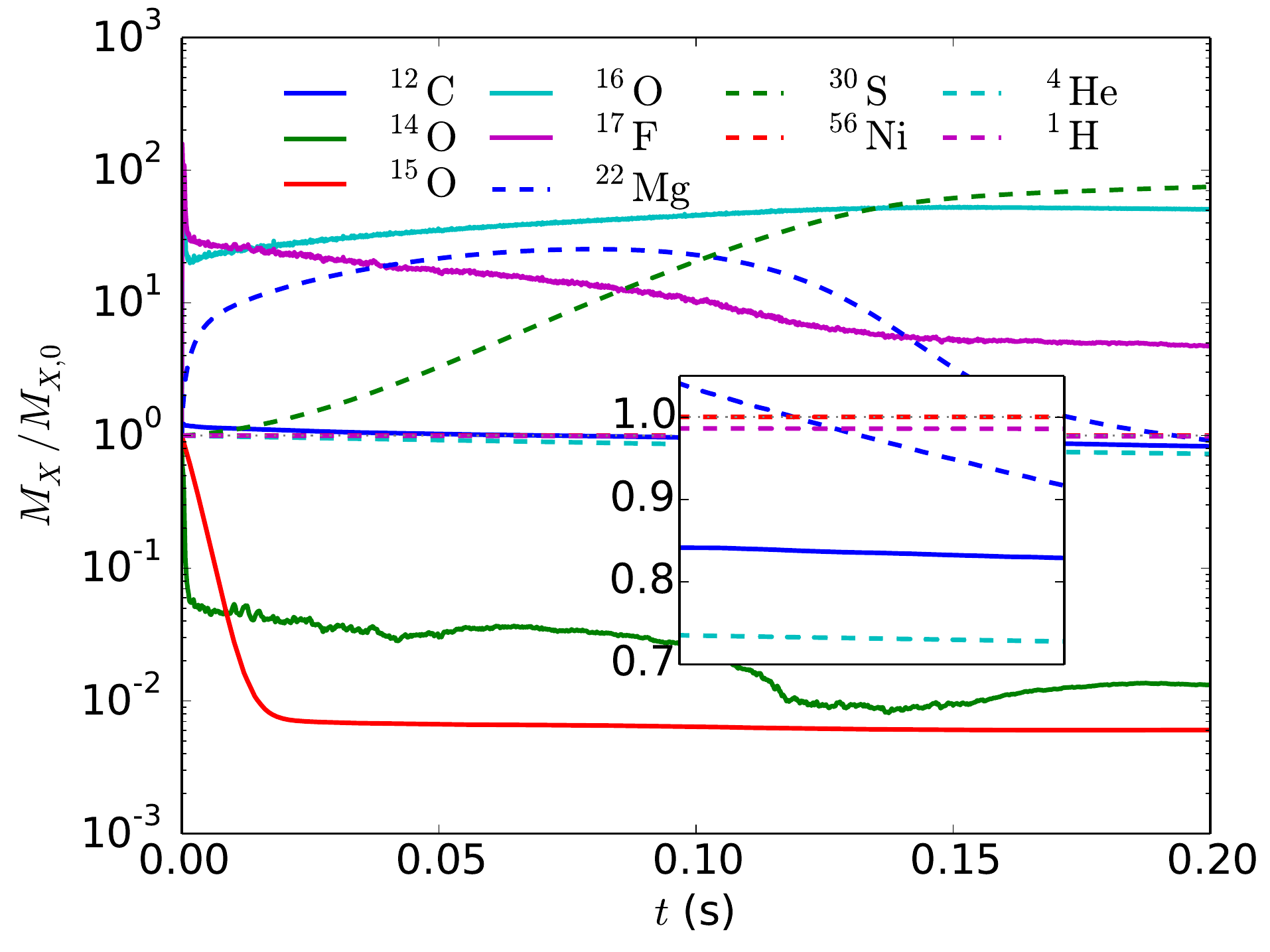}
  \caption{\label{fig:masses} Evolution of the total mass for each
    species normalized to its starting value.  The inset plot shows a
    zoom in on the last 10 ms of evolution and uses a linear scale.
    \isot{O}{16} and \isot{F}{17} were out of equilibrium due to
    approximations made in constructing the initial model.  The total
    production of \isot{Ni}{56} from \isot{S}{30} is negligible.  The
    total amount of \isot{He}{4} and \isot{H}{1} burned by mass was
    $\lesssim 27\%$ and $\lesssim 1.5\%$, respectively.}
\end{figure}

Very little \isot{S}{30} was burned to \isot{Ni}{56}.  To disentangle
the amount of dredge up from the amount produced from reactions, it is
instructive to look at the total mass of each species.  Figure
\ref{fig:masses} shows the total mass of each isotope in our network,
normalized to that isotope's initial total mass, as a function of
time; the inset plot shows a zoom in on the last 10 ms of evolution.
There were a few isotopes---e.g.~ \isot{F}{17} and \isot{O}{16}---in
our initial model that were out of equilibrium because of our
selecting only the most abundant species from the Kepler models and
then setting the abundance of the rest of the species to zero.  These
species were very quickly produced in a transient event at the
beginning of the simulation.  The \isot{F}{17} produced in this
transient is burned away to \isot{Mg}{22} somewhat quickly initially,
but is also replenished by \isot{O}{14}\ap\isot{F}{17}; the
competition of these two reaction chains results in a slowly
decreasing \isot{F}{17} abundance.  Likewise, the \isot{O}{14}
converted to \isot{F}{17} is quickly replenished by burning
\isot{C}{12}.  Indeed, the \isot{C}{12}\pg\isot{N}{13} reaction rate
is the most prodigious rate during the simulation, and any
\isot{C}{12} is nearly instantaneously converted to \isot{O}{14}.  The
total amount of \isot{Mg}{22} peaks around $t\sim0.09$ s, even though
a significant amount of \isot{S}{30} has been produced.  The total
amount of \isot{He}{4} and \isot{H}{1} burned by mass was $\lesssim
27\%$ and $\lesssim 1.5\%$, respectively.  A trace amount ($< 10^{-5}$
of the total mass) of material was lost off the grid due to expansion
of the atmosphere.

In absolute terms, the total increase in \isot{Ni}{56} mass in the
atmosphere ($r \geq 1500$ cm) --- accounting for mass lost from the
grid --- was $\sim3.99\times10^{11}$ g.  The total production of
\isot{Ni}{56} from burning was $\sim2.24\times10^{10}$ g.  The nuclear
production of \isot{Ni}{56} was confined to the atmosphere, and,
therefore, the total amount of \isot{Ni}{56} dredge up during the
simulation is the difference between the above masses, or
$\sim3.77\times10^{11}$ g.  Alternatively, in our simulations
approximately 94\% of the increase of \isot{Ni}{56} in the atmosphere
is due to dredge up from convective overshoot.  We note that the
severity of convective overshoot depends on both the resolution and
dimensionality of the simulation.  As in Paper I, we note that
increasing resolution generally tends to decrease the amount of
overshoot until some convergence is reached.  We suspect that a
three-dimensional simulation will also show reduced overshoot compared
to studies presented here; we leave the three-dimensional studies for
a future paper.

\section{Discussion and Conclusions}\label{Sec:Conclusion}

We presented the first multidimensional calculations of convective
flow in the context of mixed H/He X-ray bursts using realistic
hydrodynamics and an approximate reaction network for H/He-burning.
We demonstrated convergence of our results and explored the behavior
of the nucleosynthesis and convection.  The approximate network we
utilized assumed some reaction chains occurred faster than what would
happen in a larger network.

In addition, the approximate network utilized here resulted in the
developement of two layers of convection as a result of the existence
of a secondary peak in energy gernation rate, which eventually
dominated the total energy output.  We traced the cause of this
secondary peak in energy generation rate to the critical rp-process
breakout branching ratio, $\lambda_1$, which measures the ratio of the
$\beta^+$-decay rate versus $\alpha$-capture rate of \isot{Ne}{18}.
With our particular initial conditions, this branching ratio sharply
transitions between zero and one midway through our atmosphere.  

The secondary peak in energy generation rate becomes comparable to the
primary peak around $t\sim0.1$ s (see Figure \ref{fig:enuclate}).  The
split of the convective region lags behind the growth of the energy
generation rate peak.  Indeed, the two layered structure is not well
defined until about $t\sim0.14$ s.  A small entropy jump between the
two layers prevents material from crossing between each layer.  This
causes a quenching of the mixing of isotopes, and alters the burning.
Coincidentally, Figure \ref{fig:resstudy} shows that the rate of
increase of the peak temperature changes around the time of layer
formation.

We note that we artificially boosted the temperature and density at
the base of our accreted layer to give prodigous burning (see the
discussion in \ref{Sec:Initial Model}).  Simple Kepler calculations
show that in order to reproduce a model with characteristics similar
to our base density, the initial metallicity of the accreted material
has to be turned down significantly, $Z\sim 10^{-4}$, making such a
system perhaps rare in nature.  Continuing these calculations in
one-dimension, the full network in Kepler showed a very slight
non-monotonicity of the energy generation rate in a region containing
composition profiles qualitatively similar to what we see near the
development of our secondary peak in energy generation rate.  However,
this minor bump in Kepler's energy generation rate was transient and
certainly never dominated the burning as we see in our calculations
with the approximate network.  The development of layered convective
regions in our simulations is likely artificial and due to the
approximations used in the reaction network.

These calculations are a starting point for more realistic
calculations of H/He X-ray bursts.  Future work will focus on
improving the nuclear physics, moving to three dimensions, and
considering larger domains.  We note that very little hydrogen and
helium were processed during the short duration of our
simulations---and similar Kepler calculations show very little change
in energy generation rate profiles on this timescale.  This suggests
that even further (and possibly more accurate) simplifications can be
made to the network for use on the short timescales amenable to
multidimensional simulations.  Indeed, the original approximate
network of \citet{wallacewoosley:1981} was designed (in part) to model
the full burst cycle where copious amounts of heavy elements were
produced.  Improvements to the nuclear physics and approximations
therein are work for a future paper.

In addition, one path that will enable larger simulations is to
develop a subgrid model for the burning, and using more realistic
initial models (perhaps following the methodology from
\citealt{CUMMING_BILDSTEN_00}, or better mapping techniques between
Kepler models and Maestro models).  It seems that we can capture the
convective behavior on a moderate-resolution grid, but the steep
temperature dependence in the reactions requires fine resolution where
the energy generation peaks.  This was much more extreme in the case
of pure He bursts (Paper~I) than in the calculations here, however, a
potential path forward is to use subgrid resolution for the burning
and average the resulting energetics and compositions back to the
hydrodynamic grid.  Ultimately, we would like to model a laterally
propagating burning front with realistic nuclear physics.  In the
context of Maestro, this will require support for lateral gradients
instead of a single base state.  This development will be explored in
tandem with the follow-on simulations described above.

\acknowledgements 

The authors thank the referee for their useful comments that helped
clarify some points of the paper.  We especially thank Stan Woosley
for providing us with data from simple Kepler burns allowing us to
compare our implementation of the approximate network to a full
production network.  We also thank Alex Heger for sharing some Kepler
X-ray burst models, upon which we based the parameterized models used
here.  As always, we thank Frank Timmes for making his equation of
state publicly available.  The work at Stony Brook was supported by
DOE/Office of Nuclear Physics grants No.~DE-FG02-06ER41448 and
DE-FG02-87ER40317 to Stony Brook.  This work was also supported, at
UCSC, by the National Science Foundation (AST 0909129), the NASA
Theory Program (NNX09AK36G), and especially by the DOE HEP Program
through grant DE-FC02-06ER41438.  The work at LBNL was supported by
the Applied Mathematics Program of the DOE Office of Advance
Scientific Computing Research under U.S. Department of Energy under
contract No.~DE-AC02-05CH11231.  This research used resources of the
National Energy Research Scientific Computing Center, which is
supported by the Office of Science of the U.S. Department of Energy
under Contract No. DE-AC02-05CH11231.

\clearpage

\appendix

\section{\label{app:models} Initial Models}

We define our parameterized models to consist of three regions: an
isothermal lower region, representing the underlying neutron star, a ramp-up
region to bring us up to the temperature at the base of the accreted
layer, and an isentropic region at the surface which is where the
convection will take place.

We specify the temperature in the lower isothermal region, $T_\mathrm{ns}$,
the temperature at the base of the accreted layer, \Tb,
the density at the base of the accreted layer, \rhob,
the width of the transition region, $\delta$, and the lowest temperature
in the convection region, $T_\mathrm{cutoff}$.  Additionally, we specify
the composition in the accreted layer, $X_\mathrm{fuel}$, and the composition
in the underlying neutron star, $X_\mathrm{ns}$.  For all these calculations,
we hold the gravitational acceleration, $g$, constant.  The basic procedure
follows the methodology outlined in \citet{ppm-hse}.

We set the start the base of the accreted layer a distance $H$ from
the bottom of the domain.  The temperature and composition are set to
be:
\begin{eqnarray}
X_i &=& X_\mathrm{fuel} + \frac{1}{2} (X_\mathrm{fuel} - X_\mathrm{ns})
        \left [ 1 + \tanh \left (\frac{r_i - H}{\delta} \right ) \right ] \\
T_i &=&  T_\mathrm{base} + \frac{1}{2} (T_\mathrm{base} - T_\mathrm{ns})
        \left [ 1 + \tanh \left (\frac{r_i - H}{\delta} \right ) \right ] 
\end{eqnarray}
This temperature serves as an initial guess in the isentropic region
and will be reset there.

We choose the base of the
accreted layer as the starting point of integration.  We determine 
the base pressure and entropy from the equation of state:
\begin{equation}
p_\mathrm{base} = p(\rho_\mathrm{base}, T_\mathrm{base}, X_\mathrm{fuel}); \qquad
s_\mathrm{base} = s(\rho_\mathrm{base}, T_\mathrm{base}, X_\mathrm{fuel})
\end{equation}
We integrate radially outward from the base, using the condition of
hydrostatic equilibrium to find the needed pressure while constraining
the thermodynamic conditions to yield constant entropy.  Finding the
density, $\rho_i$, and temperature, $T_i$, in zone $i$ is an iterative
procedure:
\begin{itemize}
\item pick an initial guess for $\rho_i$ and $T_i$

\item compute the hydrostatic pressure:
\begin{equation}
p_i^\mathrm{HSE} = p_{i-1} + \frac{1}{2} \Delta r (\rho_i + \rho_{i-1}) g
\end{equation}

\item define
\begin{eqnarray}
F &\equiv& p_i^\mathrm{HSE} - p_i^\mathrm{EOS}  \\
G &\equiv& s_\mathrm{base} - s_i^\mathrm{EOS}  
\end{eqnarray}
where $p_i^\mathrm{EOS} = p(\rho_i, T_i, X_\mathrm{fuel})$ and
$s_i^\mathrm{EOS} = s(\rho_i, T_i, X_\mathrm{fuel})$ from the equation
of state.  We then use a Newton-Raphson method to correct $\rho_i$ and
$T_i$ subject to $F = G = 0$.  

\item we continue to iterate on this zone util we converge.  If the
 temperature falls below $T_\mathrm{cutoff}$, then we switch to 
 constraining only to the hydrostatic pressure, keeping the temperature
 constant.  

\end{itemize}

This defines the isentropic region above the base of the accreted layer.
Beneath the base (including the transition region characterized by width $\delta$),
we integrate radially inward, constraining the pressure to
that demanded by HSE:
\begin{equation}
p_i^\mathrm{HSE} = p_{i+1} - \frac{1}{2} \Delta r (\rho_i + \rho_{i+1}) g
\end{equation}
Since we are using the prescribed temperature, we only need to zero
$F \equiv p_i^\mathrm{HSE} - p_i^\mathrm{EOS}$ using a Newton-Raphson 
technique.

The code to generate these initial models is provided in the Maestro
source release in {\tt MAESTRO/Util/initial\_models/toy\_atm}.

\clearpage

\clearpage

\begin{deluxetable}{cc}
\tablecolumns{2}
\tablewidth{0pt}
\tablecaption{\label{table:rates} ReacLib Reaction Rates Used in Approximate Network}
\tablehead{%
  \colhead{Reaction} &
  \colhead{ReacLib Reference Label}}
\startdata
Triple--$\alpha$ & fy05 \\
\isot{C}{12}\pg\isot{N}{13} & ls09 \\
\isot{N}{14}\pg\isot{O}{15} & im05 \\
\isot{O}{14}\weak\isot{N}{14} & wc07 \\
\isot{O}{14}\ap\isot{F}{17} & Ha96c \\
\isot{O}{15}\weak\isot{N}{15} & wc07 \\
\isot{O}{15}\ag\isot{Ne}{19} & dc11 \\
\isot{O}{16}\ag\isot{Ne}{20} & co10 \\
\isot{O}{16}\pg\isot{F}{17} & ia08 \\
\isot{F}{17}\weak\isot{O}{17} & wc07 \\
\isot{F}{17}\gp\isot{O}{16} & ia08 \\
\isot{F}{17}\pg\isot{Ne}{18} & cb09 \\
\isot{Ne}{18}\weak\isot{F}{18} & wc07 \\
\isot{Ne}{18}\ap\isot{Na}{21} & mv09 \\
\isot{Ne}{19}\weak\isot{F}{19} & wc07 \\
\isot{Ne}{19}\pg\isot{Na}{20} & il10 \\
\isot{Si}{26}\ap\isot{P}{29} & ths8 \\
\isot{Ti}{44}\ap\isot{V}{47} & rh10
\enddata
\end{deluxetable}

\clearpage

\begin{deluxetable}{cc}
\tablecolumns{2}
\tablewidth{0pt}
\tablecaption{\label{table:model_params} Initial model parameters}
\tablehead{parameter & value}
\startdata
$\rhob$             & $2\times 10^6~\gcc$ \\
$T_\mathrm{ns}$     & $3\times 10^8$~K \\
$\Tb$               & $9.5\times10^8$~K \\
$T_\mathrm{cutoff}$ & $5\times 10^7$~K \\
$H$                 & $1450$~cm \\ 
$\delta$            & $12.0$~cm \\
$g$                 & $-2.45\times 10^{14}~\mathrm{cm~s^{-2}}$ \\
\hline
\multicolumn{2}{c}{fuel composition, $X_\mathrm{fuel}$} \\
\hline
\isot{H}{1}         & 0.72 \\ 
\isot{He}{4}        & 0.24 \\
\isot{O}{14}        & $4\times 10^{-4}$ \\
\isot{O}{15}        & $3\times 10^{-3}$ \\
\isot{Mg}{22}       & $10^{-3}$ \\
\isot{S}{30}        & $10^{-3}$ \\
\isot{Ni}{56}       & $0.0346$ \\
\hline
\multicolumn{2}{c}{neutron star composition, $X_\mathrm{ns}$} \\
\hline
\isot{Ni}{56}       & 1.0  \\
\enddata
\end{deluxetable}  

\clearpage

\clearpage

\clearpage

\clearpage

\clearpage

\clearpage

\clearpage

\clearpage

\clearpage

\clearpage

\clearpage

\clearpage

\end{document}